\documentclass[journal]{IEEEtran}
\IEEEoverridecommandlockouts
\def\linspread{1}
\def\linspreadalgr{1}
\usepackage{tablefootnote}
\usepackage{cite}
\usepackage{color}
\usepackage{amsfonts}
\usepackage{amssymb}
\usepackage{amsmath}
\usepackage{amsthm}
\usepackage{bm}
\usepackage{algorithm}
\usepackage{algorithmic}
\usepackage{array}
\usepackage{booktabs} 
\usepackage{float} 
\usepackage{multirow}
\usepackage{balance}  
\usepackage{graphicx}
\usepackage{subcaption}  
\usepackage{textcomp}
\usepackage{stfloats}  
\usepackage{url}
\usepackage{verbatim} 
\usepackage{graphicx}
\usepackage{epstopdf}
\usepackage{lettrine}
\usepackage{balance}  
\usepackage[colorlinks=false,
linkcolor=blue,
anchorcolor=blue,
citecolor=blue]{hyperref}
\graphicspath{ {figs/system-model/} {figs/simulation/} {figs/biography/} }

\DeclareMathOperator*{\argmin}{arg\,min}
\theoremstyle{definition}
\newtheorem{theorem}{Theorem}
\newtheorem{proposition}[theorem]{Proposition}

\theoremstyle{remark}

\newtheorem{lemma}{\indent Lemma}

\allowdisplaybreaks[3]
\usepackage{afterpage}

\def\BibTeX{{\rm B\kern-.05em{\sc i\kern-.025em b}\kern-.08em
		T\kern-.1667em\lower.7ex\hbox{E}\kern-.125emX}}

\makeatletter

\newcommand{\Rmnum}[1]{\expandafter\@slowromancap\romannumeral #1@}
\makeatother

\newfloat{figtab}{thb}{fgtb}
\makeatletter
\newcommand\figcaption{\def\@captype{figure}\caption}
\newcommand\tabcaption{\def\@captype{table}\caption}
\makeatother

\begin{document}

\title{Hybrid-Field Joint Channel and Visible Region Estimation for RIS-Assisted Communications
}

\author{Xiaokun Tuo, \textit{Graduate Student Member, IEEE},  Ming-Min Zhao, \textit{Senior Member, IEEE}, Xiang Wang, Changsheng You, \textit{Member, IEEE}, and Min-Jian Zhao, \textit{Member, IEEE}
\thanks{X. Tuo, M. M. Zhao and M. J. Zhao  are with the College of Information Science and Electronic Engineering, Zhejiang University, Hangzhou 310027, China and also with the Zhejiang Provincial Key Laboratory of Multi-Modal Communication Networks and Intelligent Information Processing, Hangzhou 310027, China (e-mail: xktuo@zju.edu.cn; zmmblack@zju.edu.cn; mjzhao@zju.edu.cn).

X. Wang is with the School of Information and Navigation, Air Force Engineering University, Xi'an 710077, China (e-mail: lleafwx626@126.com).

C. You is with the Department of Electronic and Electrical Engineering, Southern University of Science and Technology (SUSTech), Shenzhen 518055, China (e-mail: youcs@sustech.edu.cn).
		}}
        
\maketitle

\begin{abstract}
In reconfigurable intelligent surface (RIS)-assisted millimeter-wave (mmWave) communication systems, the large-scale RIS introduces pronounced geometric effects that lead to the coexistence of far-field and near-field propagation. Furthermore, random blockages induce spatial non-stationarity across the RIS array, causing signals from different scatterers to illuminate only partial regions, referred to as visible regions (VRs). This renders conventional far-field and fully visible array-based channel models inadequate and makes channel estimation particularly challenging. In this paper, we investigate the non-stationary cascaded channel estimation problem in a hybrid-field propagation environment, where the RIS-base station (BS) link operates in the far-field, while the user-RIS link exhibits near-field characteristics with partial visibility. To address the resulting high-dimensional and coupled estimation problem, a reduced-dimensional sparse bilinear representation is developed by exploiting the structural characteristics of the cascaded channel. In particular, a dictionary compression technique is proposed to represent the high-dimensional coupled dictionary using a low-dimensional polar-domain dictionary weighted by a visibility matrix, thereby significantly reducing the problem scale. Based on this representation, a turbo-structured joint Bayesian estimation (TS-JBE) approach is proposed to simultaneously estimate the channel gains, VRs, and off-grid parameters, thereby avoiding error propagation inherent in existing sequential methods. Simulation results demonstrate that the proposed method significantly improves the estimation accuracy compared with existing approaches.
\end{abstract}


\vspace{-0.0cm}

\section{Introduction}
\label{sec:Introduction}

\lettrine[lines=2]{W}{ith} the rapid increase in the number of mobile devices and the proliferation of wireless applications, future wireless networks are facing unprecedented challenges in terms of peak data rate, spectral efficiency, energy efficiency, and reliability \cite{Lu2024NF,you20256G}. As a promising candidate for sixth-generation (6G) communications, reconfigurable intelligent surfaces (RISs) have recently attracted significant attention due to the capability of reconfiguring wireless propagation environment in a cost-effective and energy-efficient manner \cite{shao2024ris,ri2024ris}.
A RIS typically consists of a large number of nearly passive reflecting elements whose reflection coefficients can be programmed to manipulate the phase of incident electromagnetic waves \cite{shao2024ris,ri2024ris}.
Through appropriate configuration of these elements, RISs can provide additional degrees of freedom to enhance signal coverage, suppress interference, and improve both spectral and energy efficiency. Owing to their low hardware cost, negligible power consumption, and ease of deployment, RISs can be densely integrated into existing wireless infrastructures, making them particularly attractive for large-scale and high-frequency 6G scenarios.

The performance gains of RIS-assisted systems, however, critically rely on the availability of accurate channel state information (CSI). Unlike conventional active antenna arrays, RIS elements are passive and incapable of signal reception or baseband processing, which makes CSI acquisition particularly challenging \cite{Zheng2022ris}. Moreover, RIS usually comprises a large number of reflecting elements, causing the dimension of RIS-associated cascaded channels to scale rapidly with the array size. Although on-off reflection schemes have been proposed to decouple the cascaded channels by selectively activating RIS elements during the training phase, thereby enabling the use of conventional channel estimation methods \cite{Mis2019onff}, such approaches still incur excessive training overhead and high computational complexity.
To address these challenges, extensive research efforts have been devoted to channel estimation in RIS-assisted communication systems. To enable channel estimation with all RIS elements simultaneously activated, methods based on conventional estimators, such as least-squares (LS) \cite{wei2022ce,you2020ls} and linear minimum mean-square error (LMMSE) \cite{li2025lmmse}, have been proposed under properly designed training protocols. In addition, compressed sensing (CS)-based approaches exploit the sparsity of cascaded channels to reduce pilot overhead, including improved orthogonal matching pursuit (OMP) algorithms \cite{bian2024omp, gao2023omp}. Moreover, matrix factorization-based methods further leverage the sparsity and low-rank properties of cascaded channels for efficient channel estimation \cite{lee2025rank, he2020matrix}.

It is worth noting that the aforementioned works are developed under far-field planar-wave assumptions. 
However, to meet the stringent performance requirements of 6G systems, RISs are expected to be deployed with a large physical aperture and operate at millimeter-wave (mmWave) or even terahertz frequencies. 
As a consequence, the Rayleigh distance increases significantly, and users or scatterers are often located in the near-field region of the RIS. 
In such scenarios, the conventional planar-wave model becomes inaccurate, and spherical-wave propagation needs to be incorporated for accurate channel modeling and estimation.
Unlike the angular-domain sparsity exhibited under the planar-wave assumption, channels modeled with spherical-wave propagation exhibit inherent sparsity in the polar domain. 
To effectively capture this property for near-field channel estimation, \cite{cui2022nearfield} proposed a novel polar-domain grid construction scheme, enabling a sparse representation of near-field channels in the polar domain. 
Building upon the spherical-wave propagation model, recent studies have explored near-field channel estimation for RIS-assisted systems. Stage-wise approaches have been developed to progressively recover the angular- and polar-domain parameters of the cascaded channel, reducing training overhead \cite{yang2023nearfield}. Beyond this, unified hybrid-field channel models have been proposed for extremely large intelligent reflecting surface (XL-IRS)-assisted orthogonal frequency division multiplexing (OFDM) systems, where tensor-based OMP combined with variational Bayesian inference (VBI) was employed to jointly estimate and track the channel \cite{tuo2025nearfield}. In addition, \cite{dong2025doubleris} considered a double-RIS-assisted mmWave communication scenario, adopted a spherical-wave-based near-field channel model, and proposed a cascaded channel modeling and analysis approach to characterize near-field propagation effects and evaluate system performance.

Beyond near-field propagation, the enlargement of RIS arrays inevitably introduces spatial non-stationarity \cite{Car2020nonstation}. 
Specifically, due to geometric effects and random blockages, signals associated with different scatterers illuminate only partial regions of the RIS, which are referred to as visible regions (VRs).
Ignoring spatial non-stationarity leads to severe model mismatch and performance degradation, especially for large-aperture RIS deployments. 
To tackle this issue, several studies have investigated joint VR detection and channel estimation for RIS-assisted communication systems. 
For instance, \cite{Han2022Localization} studied joint near-field localization and channel reconstruction in XL-RIS-assisted systems. 
However, this work mainly considered the line-of-sight (LoS) path between the RIS and the user, while neglecting non-line-of-sight (NLoS) multipath components, which limits its applicability in practical environments. 
In \cite{Yu2023nf}, a hybrid channel model was introduced, together with the concept of VRs in RIS systems, and a two-stage algorithm was proposed to perform joint channel estimation and VR detection. 
Nevertheless, the inherent error propagation caused by the two-stage processing remains unavoidable and degrades the overall estimation performance.

Motivated by the above observations, this paper investigates the cascaded channel estimation problem in RIS-assisted communication systems. We propose a joint Bayesian estimation approach that enables simultaneous inference of cascaded sparse channel gains, VRs, and off-grid parameters.
The main contributions of this paper are summarized as follows:
\begin{itemize}
  \item A hybrid-field cascaded channel model is developed for RIS-assisted mmWave systems, where the RIS-base station (BS) channel is modeled in the far-field, while the user-BS channel is characterized by near-field propagation with explicit VR modeling to capture spatial non-stationarity.
  \item By exploiting structural characteristics of the cascaded channel, a compression strategy for high-dimensional coupled dictionaries is introduced to construct a reduced-dimensional sparse bilinear channel representation, thereby significantly reducing the problem scale and effectively alleviating the resulting multi-dimensional, strongly coupled estimation problem.
  \item Based on the proposed channel representation, a turbo-structured joint Bayesian estimation (TS-JBE) approach is developed to jointly estimate the cascaded sparse channel gains, VRs, and off-grid parameters, thereby avoiding the error propagation inherent in conventional stage-wise estimation schemes.
  \item Simulation results demonstrate that the proposed approach achieves superior normalized mean squared error (NMSE) performance compared with existing approaches, validating its effectiveness under hybrid-field and spatially non-stationary propagation conditions.
\end{itemize}

The rest of this paper is organized as follows. 
Section~\ref{System Model} introduces the system model.
Section~\ref{sec:sparse_representation_Prior_Model} presents the sparse representation of the cascaded channel and the hierarchical sparse prior model.
Section~\ref{sec:TS-JBE} elaborates on the proposed TS-JBE algorithm.
Section~\ref{sec:simulation} presents numerical results to validate the effectiveness of the proposed approach.
Finally, Section~\ref{sec:conclusion} concludes the paper.

\par \textit{Notations:} Scalars, vectors and matrices are respectively denoted by lower/upper case, boldface lowercase and boldface uppercase letters. The transpose, conjugate, and conjugate transpose of a general vector $\mathbf{z}$
are denoted by $\mathbf{z}^T$, $\mathbf{z}^*$, and $\mathbf{z}^H$, respectively. The operators $\|\cdot\|$, and $|\cdot|$ denote $\ell_2$-norm, and absolute value, respectively.
The notation $\langle f(z) \rangle_{q(z)} = \int f(z) q(z)\, \mathrm{d}z$ denotes the expectation of $f(z)$ with respect to the random variable $z$ distributed according to $q(z)$.
$\mathrm{diag}(\mathbf{z})$ denotes a diagonal matrix with diagonal $\mathbf{z}$, and $\mathrm{vec}(\mathbf{Z})$ stacks the columns of $\mathbf{Z}$ into a vector. Its inverse operator $\mathrm{unvec}_{r\times c}(\cdot)$ reshapes a vector of length $rc$ into an $r\times c$ matrix in a column-wise manner. $\Re(\cdot)$ and $\Im(\cdot)$ denote the real and imaginary parts, respectively.
$\odot$, $\otimes$, $\ast$, and $\bullet$ denote the Hadamard product, Kronecker product, row-wise Khatri-Rao product, and column-wise Khatri-Rao product, respectively. $\mathbb{R}$ and $\mathbb{C}$ denote the sets of real and complex numbers.
$\mathcal{CN}(\boldsymbol{z}; \boldsymbol{\mu}, \boldsymbol{\Sigma})$ denotes the
complex Gaussian distribution with mean $\boldsymbol{\mu}$ and covariance
$\boldsymbol{\Sigma}$, and $\Gamma(z; a, b)$ denotes the Gamma distribution
with shape parameter $a$ and rate parameter $b$.

\section{System Model}
\label{System Model}
\begin{figure}[t]
    \centering
    \includegraphics[width=0.45\textwidth, height=0.25\textwidth, keepaspectratio]{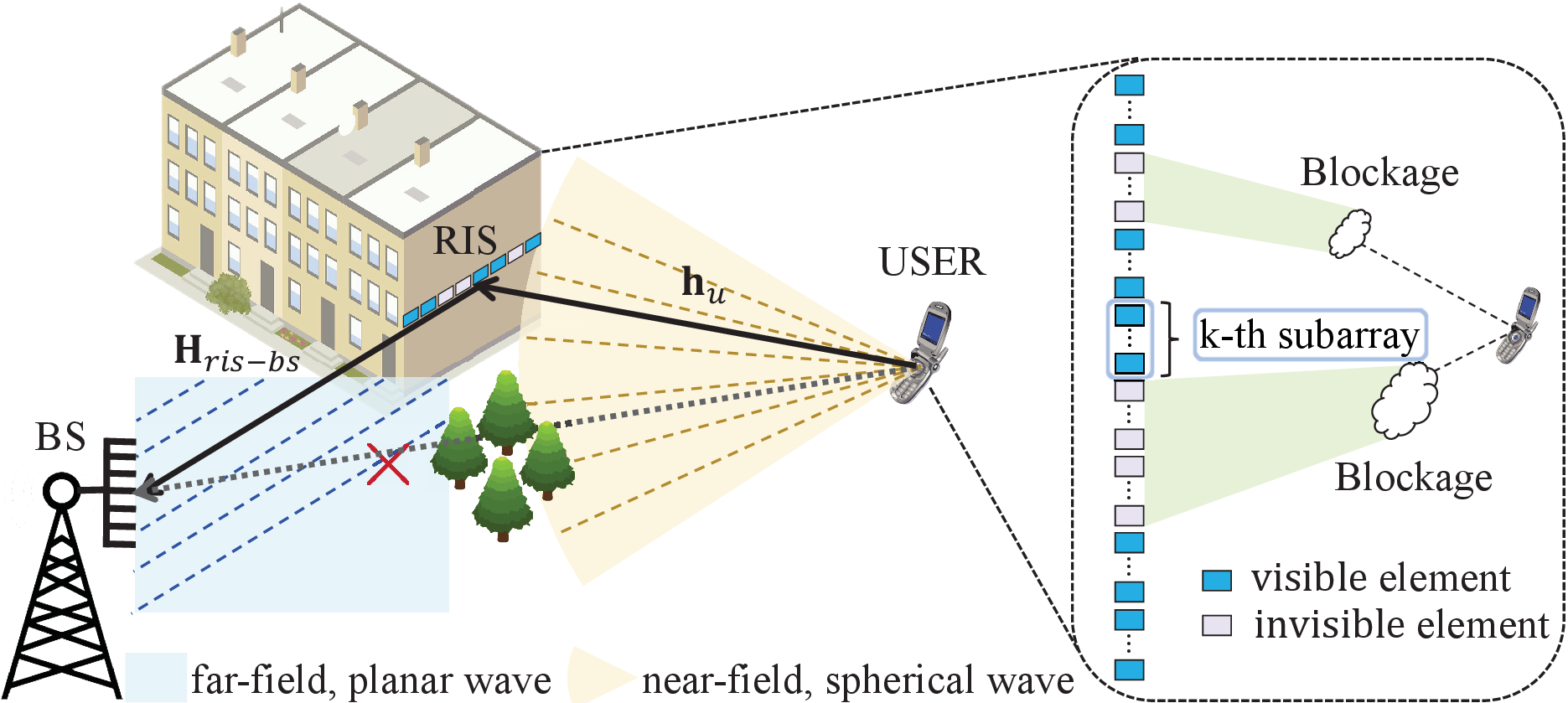}
    \caption{Considered RIS-aided mmWave system model.}
    \label{pic:sys_mod}
\end{figure} 
As illustrated in Fig. \ref{pic:sys_mod}, consider a RIS-aided multiple-input multiple-output (MIMO) system operating in the mmWave band under the time-division duplex (TDD) mode. 
The BS is equipped with $M$ antennas, while the RIS comprises $N$ reflecting elements, satisfying $N \gg M$.
The RIS is partitioned into \(K\) subarrays, each consisting of \(N/K\) reflecting elements.
Within each subarray, all elements are assumed to share identical visibility, since the channel can be regarded as spatially stationary over a small aperture~\cite{yuan2023k}.
A single-antenna user is served, with the RIS deployed in its vicinity to enhance the communication performance.\footnote{The proposed approach can be extended to the multiuser scenario by allowing each user to transmit orthogonal pilot sequences, thereby enabling the separation and estimation of the cascaded channels for each user.}
Both the BS and RIS employ uniform linear arrays (ULAs) with half-wavelength spacing, i.e., $d=\lambda/2$. 
The direct user-BS link is assumed to be blocked by obstacles (e.g., trees or buildings), so that communication relies solely on the cascaded user-RIS and RIS-BS channels, with $\mathbf{h}_{u}\in\mathbb{C}^{N\times 1}$ and $\mathbf{H}\in\mathbb{C}^{M\times N}$ denoting the corresponding channel gains, respectively.

\subsection{Channel Model}
\label{channel Model}
In conventional array modeling, targets located in the far-field region are typically characterized using the planar-wave assumption. 
The array response vector (ARV) of an $N$-element ULA is then $ \mathbf{a}^{f}(\varphi) = [1, e^{-j\frac{2\pi}{\lambda} d \varphi}, \cdots, e^{-j\frac{2\pi}{\lambda} (N-1) d \varphi}]^T$, where $d$ is the inter-element spacing, $\lambda$ is the carrier wavelength, and $\varphi = \cos\phi$ is the spatial angle corresponding to the physical angle of arrival (AoA) or departure (AoD).
When the target enters the near-field region, the planar-wave model is no longer valid, and a spherical-wave assumption is required to characterize the wavefront curvature \cite{Lu2024NF}. Accordingly, the near-field ARV can be expressed as $
\mathbf{a}^{n}(\vartheta, r)
= [e^{-j\frac{2\pi}{\lambda} (r^{(1)}(\vartheta,r) - r)}, \!\cdots\!, e^{-j\frac{2\pi}{\lambda} (r^{(N)}(\vartheta,r) - r)}]^T
$, where $\vartheta = \cos\theta$ denotes the spatial angle, $r$ denotes the distance from the reference element to the target, and $r^{(n)}(\vartheta,r)$ denotes the
distance between the $n$-th array element and the target, which is uniquely
determined by $(\vartheta,r)$ and the array geometry.
The near- and far-field regions are separated by the Rayleigh distance $Z = \frac{2(D_r^2 + D_s^2)}{\lambda}$, 
where $D_r$ and $D_s$ denote the receive and transmit array apertures\cite{Lu2024NF}.
In mmWave RIS-aided MIMO systems, the large number of antennas or reflecting elements results in an enlarged physical aperture.
Since the RIS is typically deployed in close proximity to the user, the user-RIS distance is often comparable to or even smaller than the Rayleigh distance. Consequently, the user-RIS link operates in the near-field region and should be modeled using a spherical-wave propagation assumption \cite{liu2021near}.
Accordingly, the user-RIS and RIS-BS links are modeled using spherical-wave and planar-wave propagation, respectively.
Without loss of generality, we consider a Cartesian coordinate system in which the RIS is placed along the $x$-axis, with its leftmost element located at the origin, i.e., $\mathbf{n} = [0, 0]^T$. The BS array is positioned along the $y$-axis, and the topmost BS element is located at $\mathbf{m} = [x_{BS}, y_{BS}]^T$.\footnote{Although a 2D coordinate system is considered here for simplicity, the proposed method can be directly extended to a 3D coordinate system by appropriately modifying the array response dictionaries.} Based on this geometric configuration, the cascaded channel can be modeled as follows.

\subsubsection{User-RIS Channel}
For the $l$-th propagation path between the user and RIS, where 
$l \in \mathcal{L}_U \triangleq \{1,\ldots,L_U\}$ and $L_U$ denotes the number of user-related paths ($l=1$ corresponding to the LoS path from the user), 
the position of the corresponding scatterer (or the user) is given by $\mathbf{r}_{U,l} = 
\big[ r_{U,l} \cos\theta_{U,l},\, r_{U,l} \sin\theta_{U,l} \big]^{T}$, where $\theta_{U,l}$ denotes the physical AoA of the $l$-th path, and $r_{U,l}$ represents the reference distance from the RIS array to the corresponding scatterer (or the user). In particular, $\mathbf{r}_{U,1} = \mathbf{u}$ corresponds to the user location. Accordingly, the distance between the $n$-th RIS element and the $l$-th scatterer (or the user) is given by $r_{U,l}^{(n)} = \| \mathbf{r}_{U,l} - \mathbf{n}_n \|$, where $\mathbf{n}_n = [(n-1)d,\,0]^{T}$ denotes the position vector of the $n$-th RIS element. Based on the spherical-wave propagation model, the near-field ARV
associated with the $l$-th path is defined as $
\mathbf{a}(\vartheta_{U,l}, r_{U,l}) 
= \big[ e^{-j\frac{2\pi}{\lambda}(r_{U,l}^{(1)} - r_{U,l})}, \ldots, 
e^{-j\frac{2\pi}{\lambda}(r_{U,l}^{(N)} - r_{U,l})} \big]^{T}
$, where $\vartheta_{U,l} = \cos\theta_{U,l}$ denotes the corresponding spatial angle.
Using the Fresnel approximation \cite{Selvan2017fraun}, the distance $r_{U,l}^{(n)}$ can be approximated as
\begin{align}
\label{Fresnel approximation}
r_{U,l}^{(n)} 
&\approx r_{U,l} - (n-1)d \vartheta_{U,l} 
+ \frac{d^2 (n-1)^2 (1 - \vartheta_{U,l}^2)}{2 r_{U,l}}.
\end{align}
Furthermore, a binary indicator vector is usually employed to capture the non-stationary characteristics of the channel \cite{Zhou2024VR,Tang2024VR,Han2022Localization}. 
We introduce $\boldsymbol{\phi}_l \in \{0,1\}^{N \times 1}$ to indicate the visibility of RIS elements for the $l$-th propagation path. 
Specifically, the $n$-th entry of $\boldsymbol{\phi}_l$ is equal to $1$ if the $n$-th RIS element is visible to the $l$-th path, and $0$ otherwise. With this definition, the user-RIS channel can be modeled as
\begin{equation}
\mathbf{h}_u = \sum_{l=1}^{L_U} 
\alpha_{l} \, \mathbf{a}(\vartheta_{U,l}, r_{U,l}) \odot \boldsymbol{\phi}_{l},
\label{eq:h_u}
\end{equation}
where $\alpha_l$ denotes the complex-valued path gain of the $l$-th path.

\subsubsection{RIS-BS Channel}
For the BS-RIS MIMO channel, 
each path is characterized by an AoA at the BS and an AoD at the RIS. Specifically, the AoA and AoD of the $l$-th path are denoted by $\vartheta_{B,l} = \cos\theta_{B,l}$ and $\varphi_{R,l} = \cos\phi_{R,l}$, respectively, where $l \in \mathcal{L}_{RB} \triangleq \{1, \ldots, L_{RB}\}$ and $L_{RB}$ denotes the total number of propagation paths between the RIS and BS. 
Accordingly, the RIS-BS channel can be expressed as
\begin{equation}
\mathbf{H} = \sum_{l=1}^{L_{RB}} \beta_l \, 
\mathbf{a}_{B}(\vartheta_{B,l}) \mathbf{a}_{R}^H(\varphi_{R,l}),
\end{equation}
where $\beta_l$ denotes the complex-valued gain of the $l$-th path.

\subsubsection{Hybrid-field Cascaded Channel}
The effective channel from the user to the BS via the RIS is determined by the RIS-BS channel $\mathbf{H}$, 
the user-RIS channel $\mathbf{h}_u$, and the configurable phase shifts applied at the RIS. 
Specifically, the $n$-th RIS element imposes a phase shift $\nu_n \in [0,2\pi)$ on the reflected signal, which corresponds to a reflection coefficient $\eta_n = e^{j\nu_n}$. 
Collecting all reflection coefficients yields the phase-shift vector 
$\boldsymbol{\eta} = [\eta_1, \eta_2, \ldots, \eta_N]^{T}$, 
and the associated diagonal phase-shift matrix is given by 
$\boldsymbol{\Theta} = \mathrm{diag}(\boldsymbol{\eta})$. 
Accordingly, the hybrid-field cascaded channel can be expressed as
\begin{equation}
\mathbf{h} = \mathbf{H} \, \boldsymbol{\Theta} \, \mathbf{h}_u.
\end{equation}

During uplink training, the user transmits \(P\) pilot symbols with transmit power \(P_t\).
After pilot correlation and normalization, the resulting equivalent observation is given by
\begin{equation}
\mathbf{y}_p
= \mathbf{H}\boldsymbol{\Theta}_p\mathbf{h}_u + \mathbf{n}_p
= \mathbf{h}_p + \mathbf{n}_p, \mathbf{n}_p \sim \mathcal{CN}\!(\mathbf{0}, \tfrac{\sigma^2}{P_t}\mathbf{I}_M),
\end{equation}
where \(\mathbf{h}_p \triangleq \mathbf{H}\boldsymbol{\Theta}_p\mathbf{h}_u\) denotes the effective cascaded channel associated with the \(p\)-th pilot and $\sigma^2$ denotes the noise
variance.
Stacking the signals over all pilot intervals yields
\begin{equation}
\tilde{\mathbf{y}} = [\mathbf{y}_1^T, \ldots, \mathbf{y}_P^T]^T
= \tilde{\mathbf{h}} + \tilde{\mathbf{n}},
\label{tilde_y}
\end{equation}
where \(\tilde{\mathbf{h}} = [\mathbf{h}_1^T,\ldots,\mathbf{h}_P^T]^T\) and
\(\tilde{\mathbf{n}} = [\mathbf{n}_1^T,\ldots,\mathbf{n}_P^T]^T\).

\section{Sparse Channel Representation and Prior Modeling}
\label{sec:sparse_representation_Prior_Model}
In this section, a sparse representation of the cascaded channel is developed, and the associated prior models are introduced by exploiting the structural characteristics of the RIS-aided hybrid-field propagation environment. 
Specifically, a generic pilot transmission indexed by $p$ is considered. 
For notational convenience, the pilot index $p$ is omitted whenever no ambiguity arises. 
Accordingly, the received signal, the corresponding effective cascaded channel, and the noise are denoted by 
$\mathbf{y}$, $\mathbf{h}$, and $\mathbf{n}$, respectively, each corresponding to a specific portion of $\tilde{\mathbf{y}}$, $\tilde{\mathbf{h}} $ and $\tilde{\mathbf{n}}$ indexed by $p$. 

\subsection{Sparse Representation and Model Transformation}
\label{subsec:Polar-Angle Domain Sparse Representation and Model Transformation}
\subsubsection{Sparse Representation}

To exploit the inherent sparsity of the cascaded channel, a grid-based sparse representation is adopted. Two uniform angular grids are defined for the BS and RIS as $\bm{\vartheta} \triangleq [ \vartheta_1, \cdots, \vartheta_{M} ]^T$ and $\bm{\varphi} \triangleq [ \varphi_1, \cdots, \varphi_{N} ]^T$ with $M$ and $N$ points, respectively. The grid points are given by $\vartheta_m = \frac{2}{M}\Big(m - \frac{M+1}{2}\Big)$ for $m \in \mathcal{M} \triangleq \{1,\ldots,M\}$, and $\varphi_n = \frac{2}{N}\Big(n - \frac{N+1}{2}\Big)$ for $n \in \mathcal{N} \triangleq \{1,\ldots,N\}$.
Let $\mathbf{a}_{\mathrm B}(\vartheta_m) \in \mathbb{C}^{M\times 1}$ and $\mathbf{a}_{\mathrm R}(\varphi_n) \in \mathbb{C}^{N\times 1}$ denote the far-field ARVs at the BS and RIS for the corresponding grid points. Stacking these ARVs yields the far-field array response dictionaries $\mathbf{F}_M \in \mathbb{C}^{M \!\times\! M}$ and $\mathbf{F}_N \in \mathbb{C}^{N \!\times\! N}$.
Accordingly, the RIS-BS channel admits the sparse representation $\overset{\circ}{\mathbf{H}} = \mathbf{F}_M \mathbf{A} \mathbf{F}_N^{H},
$
where $\mathbf{A} \in \mathbb{C}^{M \times N}$ is a sparse gain matrix and the superscript $(\circ)$ denotes the ideal on-grid case.

Moreover, the spatial sparsity of the near-field user-RIS channel is more naturally characterized in the polar domain. We construct a non-uniform polar-domain grid $[{\bm{\vartheta}}_u, {\mathbf{r}}_u] \triangleq [{\vartheta}_1, {r}_1; \cdots; {\vartheta}_{\bar{N}}, {r}_{\bar{N}}]$, where $\bar{N} = N_p S$, and $N_p$ and $S$ denote the numbers of angular and radial samples, respectively, following the discretization rule in \cite{cui2022nearfield}.
Let $\mathbf{a}({\vartheta}_{\bar{n}}, r_{\bar{n}})\in\mathbb{C}^{N \times 1}$ denote the near-field ARV at the $\bar{n}$-th grid point, and stack them to form the polar-domain dictionary $\mathbf{W} \in \mathbb{C}^{N \times \bar{N}}$. The user-RIS channel is then expressed as $\overset{\circ}{\mathbf{h}}_u = (\mathbf{W} \odot \boldsymbol{\Phi})\mathbf{b}$, where $\boldsymbol{\Phi}$ is a binary VR matrix and $\mathbf{b}$ is a sparse vector of complex-valued gains. 
Specifically, each nonzero $b_l$ of $\mathbf{b}$ corresponds to the $l$-th propagation path with visibility indicated by $\boldsymbol{\phi}_l$, the $l$-th column of $\boldsymbol{\Phi}$, as in~\eqref{eq:h_u}.
Based on the above representations, the overall hybrid-field cascaded channel can be rewritten as
\begin{equation}
\overset{\circ}{\mathbf{h}} = \mathbf{F}_M \mathbf{A} \mathbf{F}_N^H \, \mathrm{diag}(\boldsymbol{\eta}) (\mathbf{W} \odot \boldsymbol{\Phi}) \mathbf{b},
\label{eq:ideal_h}
\end{equation}
\subsubsection{Model Transformation}
From the cascaded channel model in \eqref{eq:ideal_h}, the unknown parameters are identified as the far-field sparse gain matrix $\mathbf{A}$, the VR matrix $\boldsymbol{\Phi}$, and the near-field sparse gain vector $\mathbf{b}$.
These parameters jointly characterize the cascaded channel via a multi-linear coupling relationship.
To facilitate efficient parameter inference, some fundamental matrix identities are exploited to reformulate \eqref{eq:ideal_h}. Specifically, by appropriately rearranging the cascaded channel expression, the received signal can be expressed in equivalent linear forms with respect to either  the VR matrix $\boldsymbol{\Phi}$ or the vectorized cascaded sparse gain $\mathrm{vec}(\mathbf{b}^T \otimes \mathbf{A})$, while treating the remaining parameters as known.
Mathematically, by defining $\bar{\mathbf{W}} = \mathbf{W} \odot \boldsymbol{\Phi}$, we have
\begin{align}
\overset{\circ}{\mathbf{h}} &= \mathbf{F}_M \mathbf{A} \mathbf{F}_N^H \, \mathrm{diag}(\boldsymbol{\eta}) (\bar{\mathbf{W}} \mathbf{b}) \nonumber \\
&= \mathbf{F}_M \mathbf{A} \mathbf{F}_N^H \, \mathrm{diag}(\bar{\mathbf{W}} \mathbf{b}) \boldsymbol{\eta} \nonumber \\
&\!\stackrel{(a)}{=} (\boldsymbol{\eta}^T \otimes \mathbf{I}_M) \, \mathrm{vec}(\mathbf{F}_M \mathbf{A} \mathbf{F}_N^H \, \mathrm{diag}(\bar{\mathbf{W}} \mathbf{b}))\nonumber \\
&\!\stackrel{(b)}{=} (\boldsymbol{\eta}^T \otimes \mathbf{I}_M) \, \mathrm{vec}(\mathbf{F}_M (\mathbf{b}^T \bar{\mathbf{W}}^T \bullet \mathbf{A} \mathbf{F}_N^H))   \nonumber \\
&\!\stackrel{(c)}{=} (\boldsymbol{\eta}^T \otimes \mathbf{I}_M) \, \mathrm{vec}(\mathbf{F}_M (\mathbf{b}^T \otimes \mathbf{A}) (\bar{\mathbf{W}}^T \bullet \mathbf{F}_N^H))   \nonumber \\
&\!\stackrel{(d)}{=} (\boldsymbol{\eta}^T \otimes \mathbf{I}_M) ((\bar{\mathbf{W}}^T \bullet \mathbf{F}_N^H)^T \otimes \mathbf{F}_M) \, \mathrm{vec}(\mathbf{b}^T \otimes \mathbf{A})\nonumber \\
&\!\stackrel{(e)}{=} (\boldsymbol{\eta}^T \otimes \mathbf{I}_M) ((\bar{\mathbf{W}} \ast \mathbf{F}_N^{\ast}) \otimes \mathbf{F}_M) \, \mathrm{vec}(\mathbf{b}^T \otimes \mathbf{A}),
\label{eq:y_x}
\end{align}
where $(a)$ and $(d)$ follow from the vectorization identity, i.e., $\mathrm{vec}(\mathbf{A}\mathbf{B}\mathbf{C}) = (\mathbf{C}^{T} \otimes \mathbf{A})\mathrm{vec}(\mathbf{B})$, 
$(b)$ holds from the property of diagonal matrices and the Khatri-Rao product, i.e., \(\mathbf{X} \, \mathrm{diag}(\mathbf{v}) = \mathbf{v}^T \bullet \mathbf{X}\),
$(c)$ follows from the property of the Khatri-Rao product, i.e., $(\mathbf{A} \otimes \mathbf{B})(\mathbf{C} \bullet \mathbf{D}) = (\mathbf{A}\mathbf{C}) \bullet (\mathbf{B}\mathbf{D})$, and $(e)$ follows from the row-wise and column-wise Khatri-Rao products relation, i.e., $\mathbf{A} \ast \mathbf{B} = \big( \mathbf{A}^T \bullet \mathbf{B}^T \big)^T$ \cite{zhang2017matrix}.

Directly estimating the cascaded sparse channel gain vector $\mathrm{vec}(\mathbf{b}^T \otimes \mathbf{A}) \in \mathbb{C}^{\bar{N}N \times 1}$
results in a prohibitively large problem size due to the direct coupling between
the near-field and far-field dictionaries, rendering practical channel estimation
computationally infeasible. Therefore, it is essential to exploit the structural characteristics of the cascaded channel model in \eqref{eq:y_x} to reduce the dimension of the estimation problem and, consequently, the computational burden.
In particular, the coupled hybrid-field dictionary term $\bar{\mathbf{W}} \ast \mathbf{F}_N^\ast$ in \eqref{eq:y_x}, arising from the interaction between the near-field and far-field dictionaries at the RIS, introduces potential column redundancy and is the primary source of the dimension expansion. If this coupled dictionary can be efficiently compressed, the overall problem dimension can be substantially reduced.
Existing studies have shown that the inner product between a near-field ARV and a dimension-compatible far-field ARV can be equivalently represented by a new near-field ARV \cite{Yu2023nf}.
This equivalence compresses the representation of the coupled dictionary from a high-dimensional angular-radial parameterization to a compact angle-distance pair.
Building on this insight, Proposition~1 further demonstrates that the row-wise Khatri-Rao product of the near-field and far-field dictionaries can be equivalently represented by a low-dimensional near-field dictionary when the near-field ARV is coupled with the VR indicator vector.
\begin{proposition}
The row-wise Khatri-Rao product between the near-field dictionary considering VR indicator $\bar{\mathbf{W}}$ and the conjugate of the far-field dictionary $\mathbf{F}_N^\ast$, i.e., $\bar{\mathbf{W}} \ast \mathbf{F}_N^{\ast}$ is equivalent to the Hadamard product between a near-field dictionary $\mathbf{Q} \in\mathbb{C}^{N\times \bar{Q}}$  with grid $[\bar{\boldsymbol{\varphi}}, \bar{\boldsymbol{r}}] \in\mathbb{R}^{\bar{Q}\times 2}$ and a VR matrix $\bar{\boldsymbol{\Phi}} \in \{0,1\}^{N \times \bar{Q}}$:
\begin{equation}
\bar{\mathbf{W}} \ast \mathbf{F}_N^\ast
\;\;\Leftrightarrow\;\;
\mathbf{Q} \odot \bar{\boldsymbol{\Phi}},
\end{equation}
where $\Leftrightarrow$ denotes that the cascaded channel $\overset{\circ}{\mathbf{h}}$
preserves the same number of active columns under the transformation between the
two dictionaries $\bar{\mathbf{W}} \ast \mathbf{F}_N^\ast$ and $\mathbf{Q} \odot \bar{\boldsymbol{\Phi}}$.
\begin{proof}
Please refer to Appendix~A.
\end{proof}
\end{proposition}
Based on Proposition~1, $\overset{\circ}{\mathbf{h}}$ can be rewritten as
\begin{align}
\overset{\circ}{\mathbf{h}}
&= (\boldsymbol{\eta}^T \otimes \mathbf{I}_M) ((\bar{\mathbf{W}} \ast \mathbf{F}_N^{\ast}) \otimes \mathbf{F}_M) \, \mathrm{vec}(\mathbf{b}^T \otimes \mathbf{A}) \nonumber \\
&=(\boldsymbol{\eta}^T \otimes \mathbf{I}_M) ((\mathbf{Q} \odot \bar{\boldsymbol{\Phi}}) \otimes \mathbf{F}_M) \, \mathrm{vec}(\mathbf{X}),
\label{eq:ideal_h_Phix}
\end{align}
where $\mathbf{X} \in \mathbb{C}^{M \times \bar{Q}}$ denotes the joint sparse matrix
associated with the transformed dictionary
$(\mathbf{Q} \odot \bar{\boldsymbol{\Phi}}) \otimes \mathbf{F}_M$.
It is worth noting that the above dictionary transformation also alters the indexing of the sparse representation.
As a result, $\mathrm{vec}(\mathbf{X})$ is generally not identical to $\mathrm{vec}(\mathbf{b}^T \otimes \mathbf{A})$, while both representations uniquely characterize the same cascaded channel. 
Moreover, the proposed transformation substantially reduces the representation dimension. Specifically, the sparse vector describing the cascaded channel is reduced from
$\mathrm{vec}(\mathbf{b}^T \otimes \mathbf{A}) \in \mathbb{C}^{M N \bar{N}  \times 1}$ 
to 
$\mathrm{vec}(\mathbf{X}) \in \mathbb{C}^{M \bar{Q}  \times 1}$, 
leading to a reduction in representation scale from $\mathcal{O}(M N^2)$ to $\mathcal{O}(M N)$, since $\bar{Q}$ is comparable to $N$, as both result from the same near-field grid generation algorithm applied to the same array.
This dimension reduction significantly alleviates the dictionary size and results in a substantial decrease in computational complexity. Building upon the reduced-dimensional representation $\mathbf{X}$, the cascaded channel can be vectorized as
\begin{equation}
\overset{\circ}{\mathbf{h}} = \mathbf{D}_1(\bar{\boldsymbol{\Phi}}) \, \boldsymbol{x},
\label{eq:ideal_h_Dx}
\end{equation}
where $\boldsymbol{x} \triangleq \mathrm{vec}(\mathbf{X})$ and
$\mathbf{D}_1(\bar{\boldsymbol{\Phi}}) = (\boldsymbol{\eta}^T \otimes \mathbf{I}_M) \, ((\mathbf{Q} \odot \bar{\boldsymbol{\Phi}}) \otimes \mathbf{F}_M)$.
This indicates that $\overset{\circ}{\mathbf{h}}$ admits a linear representation with respect to the cascaded sparse channel gain vector $\boldsymbol{x}$.

Furthermore, \eqref{eq:ideal_h_Phix} can be equivalently reformulated as
\begin{align}
\overset{\circ}{\mathbf{h}} 
&= (\boldsymbol{\eta}^T \!\otimes\! \mathbf{I}_M) ((\mathbf{Q} \!\odot\! \bar{\boldsymbol{\Phi}}) \!\otimes\! \mathbf{F}_M) \boldsymbol{x} \nonumber\\
&\!\stackrel{(f)}{=} (\boldsymbol{x}^T \!\otimes\! (\boldsymbol{\eta}^T \!\otimes\! \mathbf{I}_M))  \mathrm{vec}((\mathbf{Q} \!\odot\! \bar{\boldsymbol{\Phi}}) \!\otimes\! \mathbf{F}_M) \nonumber\\
&\!\stackrel{(g)}{=} (\boldsymbol{x}^T \!\otimes\! (\boldsymbol{\eta}^T \!\otimes\! \mathbf{I}_M))  \mathrm{vec}((\mathbf{Q} \!\otimes\! \mathbf{F}_M) \!\odot\! (\bar{\boldsymbol{\Phi}} \!\otimes\! \mathbf{1}_{M \!\times\! M})) \nonumber\\
&\!\stackrel{(h)}{=} (\boldsymbol{x}^T \!\otimes\! (\boldsymbol{\eta}^T \!\otimes\! \mathbf{I}_M))
\mathrm{diag}(\mathrm{vec}(\mathbf{Q} \!\otimes\! \mathbf{F}_M)) \mathrm{vec}(\bar{\boldsymbol{\Phi}} \!\otimes\! \mathbf{1}_{M \!\times\! M}),
\label{eq:ideal_h_vecPhi}
\end{align}
where $(f)$ follows from the vectorization identity as mentioned before, $(g)$ follows from the mixed-product property of the Kronecker and Hadamard products, i.e.,
$(\mathbf{A} \odot \mathbf{B}) \otimes (\mathbf{C} \odot \mathbf{D})
= (\mathbf{A} \otimes \mathbf{C}) \odot (\mathbf{B} \otimes \mathbf{D})$,
and $(h)$ follows from the vectorization identity of the Hadamard product, i.e.,
$\mathrm{vec}(\mathbf{A} \odot \mathbf{B})
= \mathrm{diag}(\mathrm{vec}(\mathbf{A}))\,\mathrm{vec}(\mathbf{B})$ \cite{zhang2017matrix}.

Based on the above matrix transformations, an explicit linear representation of the cascaded channel $\mathbf{h}$ with respect to $\mathrm{vec}(\bar{\boldsymbol{\Phi}} \otimes \mathbf{1}_{M \!\times\! M})$ is obtained. However, the coupling between $\mathbf{1}_{M \!\times\! M}$ and the unknown VR matrix $\bar{\boldsymbol{\Phi}}$ is undesirable, as it masks the intrinsic sparsity structure of the VR. To address this issue, Lemma~1 is further invoked to decouple $\mathrm{vec}(\bar{\boldsymbol{\Phi}} \otimes \mathbf{1}_{M \!\times\! M})$ into $\mathbf{S}(\mathbf{1}_{M \!\times\! M})\mathrm{vec}(\bar{\boldsymbol{\Phi}})$.

\begin{lemma}
\label{lemma:vectransform}
For any matrices $\mathbf{A} = [\mathbf{a}_1,...,\mathbf{a}_{c_A}] \in \mathbb{C}^{r_A \times c_A}$ and $\mathbf{B} = [\mathbf{b}_1,...,\mathbf{b}_{c_B}] \in \mathbb{C}^{r_B \times c_B}$, there exists a transformation matrix $\mathbf{S}(\mathbf{A}) \in \mathbb{C}^{r_B c_B r_A c_A  \times r_B c_B}$ such that
\begin{equation}
\label{S}
\mathrm{vec}(\mathbf{B} \otimes \mathbf{A}) = \mathbf{S}(\mathbf{A}) \, \mathrm{vec}(\mathbf{B}),
\end{equation}
where $\mathbf{S}(\mathbf{A})$ is explicitly given by
$
\mathbf{S}(\mathbf{A}) \!=\! \mathbf{I}_{c_B} \otimes 
\begin{bmatrix}
\mathbf{I}_{r_B} \!\otimes\! \mathbf{a}_1 \\
\vdots \\
\mathbf{I}_{r_B} \!\otimes\! \mathbf{a}_{c_A}
\end{bmatrix}.$
\begin{proof}
The proof is omitted due to space limitation.
\end{proof}
\end{lemma}

Using Lemma~\ref{lemma:vectransform}, the cascaded channel can be written as
\begin{equation}
\overset{\circ}{\mathbf{h}} = \mathbf{D}_2(\boldsymbol{x}) \mathrm{vec}(\bar{\boldsymbol{\Phi}}),
\label{eq:ideal_h_D_2Phi}
\end{equation}
where $\mathbf{D}_2(\boldsymbol{x}) = (\boldsymbol{x}^T \otimes (\boldsymbol{\eta}^T \otimes \mathbf{I}_M))
\mathrm{diag}(\mathrm{vec}(\mathbf{Q} \otimes \mathbf{F}_M))\mathbf{S}$ $(\mathbf{1}_{M \times M})$.
Thus, the cascaded channel $\overset{\circ}{\mathbf{h}}$ is represented as a linear model with respect to the vectorized VR matrix $\mathrm{vec}(\bar{\boldsymbol{\Phi}})$. 

Through these matrix transformations, two intuitive linear models, i.e., \eqref{eq:ideal_h_Dx} and \eqref{eq:ideal_h_D_2Phi}, are obtained. These models explicitly relate the original unknown parameters, namely, the far-field sparse matrix $\mathbf{A}$, the near-field VR matrix $\boldsymbol{\Phi}$, and the sparse gain vector $\mathbf{b}$ to the received signal through the reduced-dimensional representations $\boldsymbol{x}$ and $\bar{\boldsymbol{\Phi}}$.

\subsubsection{Off-Grid Effect Correction}
\label{off-grid}
It is important to note that the sparse representation introduced above relies on the assumption that the scatterer-, user-, and RIS-related angles or positions lie exactly on the predefined angular or polar grids. In practice, however, the spatial angles $\vartheta_{B,l}$ and $\varphi_{R,l}$, as well as the user/scatterer locations $[\vartheta_{U,l}, r_{U,l}]$, are continuous-valued, and the actual propagation parameters generally do not coincide with the discretized grids, leading to a basis mismatch. To mitigate this issue, off-grid vectors are introduced to refine the dictionary representation.

Specifically, let $m_{l} \!\triangleq\! \argmin_m | \vartheta_{B,l} \!-\! \vartheta_m |$ and $q_{l} \!\triangleq\! \argmin_q \! D( [\bar{\varphi}_l, \bar{r}_l], [\bar{\varphi}_q, \bar{r}_q])$,
for all $l \!\in\! \mathcal{L} \!\triangleq\! \{1, \!\cdots\!, L_UL_{RB}\}$,
where $m_l$ and $q_l$ denote the indices of the predefined angular and polar grid points
closest to $\vartheta_{B,l}$ and $[\bar{\varphi}_l,\bar{r}_l]$, respectively.
Here, $D(\cdot,\cdot)$ is defined as
$D([\bar{\varphi}_1,\bar{r}_1],[\bar{\varphi}_2,\bar{r}_2])
= \sqrt{\bar{r}_1^2 + \bar{r}_2^2 - 2\bar{r}_1\bar{r}_2 \cos(\bar{\phi}_2 - \bar{\phi}_1)}$,
which characterizes the distance between two polar-domain points
$(\bar{\varphi}_1,\bar{r}_1)$ and $(\bar{\varphi}_2,\bar{r}_2)$,
with $\bar{\phi}_1 = \arccos \bar{\varphi}_1$ and
$\bar{\phi}_2 = \arccos \bar{\varphi}_2$.
Then, the off-grid vectors are defined as $\Delta \bm{\vartheta} \triangleq [\Delta \vartheta_1, \cdots, \Delta \vartheta_{M}]^T$, $\Delta {\bm{\bar{\varphi}}} \!\triangleq\! [\Delta {\bar{\varphi}}_1, \!\cdots\!, \Delta {\bar{\varphi}}_{\bar{Q}}]^T$ and $\Delta {\mathbf{\bar{r}}} \!\triangleq\! [\Delta {\bar{r}}_1, \cdots, \Delta {\bar{r}}_{\bar{Q}}]^T$, where $\Delta \vartheta_m$ ($\forall m \in \mathcal{M}$), $\Delta {\bar{\varphi}}_q$ and $\Delta \bar{r}_q$ ($\forall q \in {\mathcal{\bar{Q}}}$) are given by
{\setlength\abovedisplayskip{2pt}
	\setlength\belowdisplayskip{2pt}
\begin{align}
	&\Delta \vartheta_m =  \left\{
	\begin{aligned}
		& \vartheta_{B,l} - \vartheta_m, \; \text{if}\; m = m_{l},  \\ 
		& 0,  \; \text{if}\; m \neq m_{l},
	\end{aligned}
	\right. \\
	&\Delta {\bar{\varphi}}_q = \left\{
	\begin{aligned}
		& \bar{\varphi}_l \!-\! \bar{\varphi}_q, \; \text{if}\; q \!=\! q_{l},   \\ 
		& 0,  \; \text{if}\; q \!\neq\! q_{l},
	\end{aligned}
	\right.  
	\Delta {\bar{r}}_q = \left\{
	\begin{aligned}
		&\bar{r}_l \!-\! \bar{r}_q, \; \text{if}\; q \!=\!q_{l},   \\ 
		& 0,  \; \text{if}\; q \!\neq\! q_{l},
	\end{aligned}
	\right. 	
\end{align}}%
$ \forall l \in \mathcal{L}$, respectively. 
For notational convenience, we define $\bm{\Xi} \triangleq \{ \Delta \bar{\bm{\varphi}}, \Delta \bar{\bm{r}}, \Delta \bm{\vartheta} \}$ as the off-grid parameters set. 
By incorporating $\bm{\Xi}$, the resulting channel can be expressed as
\begin{align}
\mathbf{h} 
&= (\boldsymbol{\eta}^T \otimes \mathbf{I}_M) 
   ((\mathbf{Q}(\Delta \bm{\bar{\varphi}}, \Delta {\mathbf{\bar{r}}}) \odot \bar{\boldsymbol{\Phi}}) 
   \otimes \mathbf{F}_M(\Delta \bm{\vartheta})) \, \boldsymbol{x} \nonumber \\
&= (\boldsymbol{x}^T \otimes (\boldsymbol{\eta}^T \otimes \mathbf{I}_M)) \,
   \mathrm{diag}\!(\mathrm{vec}(\mathbf{Q}(\Delta \bm{\bar{\varphi}}, \Delta {\mathbf{\bar{r}}}) 
   \otimes \mathbf{F}_M(\Delta \bm{\vartheta}))) \nonumber \\
&\quad \times \mathbf{S}(\mathbf{1}_{M \!\times\! M}) \, \mathrm{vec}(\bar{\boldsymbol{\Phi}}).
\label{eq:real_h}
\end{align}
In this way, the proposed model explicitly accounts for the off-grid deviations of the continuous propagation parameters from the predefined dictionaries, thereby mitigating basis mismatch and enabling an accurate sparse-domain representation of the cascaded channel.

\subsection{Hierarchical Sparse Prior Model}
\label{subsec: Hierarchical Sparse Probabilistic Model}
To fully exploit the sparsity and structural characteristics of the cascaded sparse channel gain vector $\boldsymbol{x}$, it is essential to adopt a structure-aware probabilistic model. Motivated by \cite{liu2020HLS}, we employ a three-layer hierarchical sparse (3LHS) prior model. Different from two-layer priors that rely solely on either the support or the variance \cite{Schniter2010Turbo,Tzi2008VBI}, the 3LHS model jointly accounts for both components, providing a flexible and robust probabilistic representation. Prior works \cite{xu2023successive,zhou2025SCVBI} have shown that this model can naturally capture structured sparsity while maintaining low computational complexity, making it particularly suitable for modeling the cascaded channel in this work. Specifically, the 3LHS model consists of three layers, i.e., the support vector $\boldsymbol{s}$, the precision vector $\boldsymbol{\rho}$, and the sparse gain vector $\boldsymbol{x}$, with the joint distribution given by
\begin{equation}
p(\boldsymbol{x}, \boldsymbol{\rho}, \boldsymbol{s})
=
p(\boldsymbol{s})\,
p(\boldsymbol{\rho}|\boldsymbol{s})
\,
p(\boldsymbol{x} | \boldsymbol{\rho}).
\label{eq:3LHS}
\end{equation}

Specifically, the support vector $\boldsymbol{s}$ is introduced to indicate whether each element of the cascaded sparse channel gain vector $\boldsymbol{x}$ is zero. In particular, $s_n = 0$ means that $x_n$ is zero, while $s_n \neq 0$ indicates that $x_n$ is nonzero. Moreover, by appropriately modeling the support vector $\boldsymbol{s}$, the unique sparse structure of $\boldsymbol{x}$ can be effectively characterized. In the considered scenario, the cascaded sparse channel gain vector $\boldsymbol{x}$ exhibits the following structural characteristics: $\mathbf{h}_{u}$ contains $L_U$ paths, while $\mathbf{H}$ contains $L_{RB}$ paths. Accordingly, the sparse matrix $\mathbf{A}$ contains $L_U$ nonzero elements, while the sparse vector $\mathbf{b}$ contains $L_{RB}$ nonzero elements. As a result, the equivalent sparse matrix $\mathbf{X}$ contains $L_U L_{RB}$ nonzero elements. Furthermore, it is readily seen that each column of $\mathbf{X}$ contains at most one nonzero element, indicating that each equivalent path on the RIS side corresponds to at most one AoD. 
This column-wise sparsity structure suggests that no explicit dependency needs to be imposed across different support entries. Therefore, modeling the support vector $\boldsymbol{s}$ as independent and identically distributed (i.i.d.) provides a sufficiently expressive yet parsimonious description of the sparsity pattern.
Consequently, for $\boldsymbol{s}$, we have
\begin{equation}
\mathrm{unvec}_{M \times \bar{Q}}(\boldsymbol{s}) =
\begin{bmatrix}
s_{1,1} & s_{1,2} & \cdots & s_{1,\bar{Q}} \\
s_{2,1} & s_{2,2} & \cdots & s_{2,\bar{Q}} \\
\vdots & \vdots & \ddots & \vdots \\
s_{M,1} & s_{M,2} & \cdots & s_{M,\bar{Q}}
\end{bmatrix} \!\in\! \{0,1\},
\end{equation}
and its probability distribution is given by
\begin{equation}
p(\boldsymbol{s}) = \prod_{m=1}^{M} \prod_{q=1}^{\bar{Q}} \left( \lambda_{m,q} \right)^{s_{m,q}} \left( 1 - \lambda_{m,q} \right)^{1 - s_{m,q}},
\end{equation}
where $\lambda_{m,q}$ is determined by the sparsity of $\boldsymbol{x}$.
Moreover, the precision vector $\boldsymbol{\rho}$ controls the variance of $\boldsymbol{x}$, e.g., $1/\rho_{m,q}$ represents the variance of ${x}_{m,q}$. The conditional probability $p(\boldsymbol{\rho} | \boldsymbol{s})$ is given by
\begin{equation}
\begin{aligned}
p(\boldsymbol{\rho} | \boldsymbol{s}) 
&= \prod_{m=1}^{M} \prod_{q=1}^{\bar{Q}} 
\Gamma(\rho_{m,q}; a_{m,q}, b_{m,q})^{s_{m,q}} \\
&\quad \times \Gamma(\rho_{m,q}; \bar{a}_{m,q}, \bar{b}_{m,q})^{1-s_{m,q}},
\end{aligned}
\end{equation}
where $\Gamma(\rho; a, b)$ serves as a conjugate prior to enable closed-form posterior inference.
Specifically, when $s_{m,q} = 1$, the inverse precision $1/\rho_{m,q}$ is expected to be close to $1$, implying that the mean of $\rho_{m,q}$ satisfies $a_{m,q}/b_{m,q} = \mathbb{E}[\rho_{m,q}] = \Theta(1)$.
Conversely, when $s_{m,q} = 0$, $1/\rho_{m,q}$ should be close to $0$, which leads to a significantly larger mean, i.e., $\bar{a}_{m,q}/\bar{b}_{m,q} = \mathbb{E}[\rho_{m,q}] \gg 1$.
Finally, based on the precision parameter $\boldsymbol{\rho}$, we model the cascaded sparse channel gain vector $\boldsymbol{x}$ with a zero-mean complex Gaussian prior, whose variance is governed by the corresponding precision parameter, i.e.,
\begin{equation}
p(\boldsymbol{x} | \boldsymbol{\rho}) = \prod_{m=1}^{M} \prod_{q=1}^{\bar{Q}} \mathcal{CN}(x_{m,q}; 0, \rho_{m,q}^{-1}).
\end{equation}

\subsection{Markov Stationary Clustering Prior Model}
\label{subsec: MS}
As mentioned in Section~\ref{System Model}, 0/1 variables are employed to indicate whether each RIS element can reflect the user’s signal, with all elements within the same subarray sharing identical visibility.  
To explicitly characterize the subarray-level visibility structure, we define the subarray VR matrix of the RIS as
$\bar{\boldsymbol{\Phi}}_s \in \{0,1\}^{K \times \bar{Q}}$.
Based on this definition, the full-RIS VR matrix is
$\bar{\boldsymbol{\Phi}} = \bar{\boldsymbol{\Phi}}_s \otimes \mathbf{1}_{\frac{N}{K} \times 1}$,
where the subarray visibility is replicated across all elements in each subarray.
Accordingly, the received signal model can be equivalently rewritten as
\begin{equation}
\mathbf{y} = \mathbf{D}_1(\bar{\boldsymbol{\Phi}}_{s},\bm{\Xi}) \boldsymbol{x} + \mathbf{n} = \mathbf{D}_2(\boldsymbol{x},\bm{\Xi}) \mathrm{vec}(\bar{\boldsymbol{\Phi}}_{s}) + \mathbf{n},    
\end{equation}
where $\mathbf{D}_1(\bar{\boldsymbol{\Phi}}_{s},\bm{\Xi}) =(\boldsymbol{\eta}^T \otimes \mathbf{I}_M) 
   ((\mathbf{Q}(\Delta \bm{\bar{\varphi}}, \Delta {\mathbf{\bar{r}}}) \odot (\bar{\boldsymbol{\Phi}}_{s} \otimes \mathbf{1}_{\frac{N}{K} \times 1})) 
   \otimes \mathbf{F}_M(\Delta \bm{\vartheta}))$ and $\mathbf{D}_2(\boldsymbol{x},\bm{\Xi}) =(\boldsymbol{x}^T \otimes (\boldsymbol{\eta}^T \otimes \mathbf{I}_M)) \,
   \mathrm{diag}\!(\mathrm{vec}(\mathbf{Q}(\Delta \bm{\bar{\varphi}}, \Delta {\mathbf{\bar{r}}}) 
   \otimes \mathbf{F}_M(\Delta \bm{\vartheta})))\mathbf{S}(\mathbf{1}_{M \!\times\! M}) \mathbf{S}(\mathbf{1}_{\frac{N}{K}} \times 1)$.
To capture the spatial correlation across subarrays, a Markov chain is introduced to characterize the dependency of visibility between neighboring subarrays \cite{Tang2024VR}. Accordingly, the prior distribution of the subarray VR matrix is given by
\begin{equation}
p(\bar{\bm{\Phi}}_s) = \prod_{q=1}^{\bar{Q}} p(v_{1,q}) \prod_{k=2}^{K} p(v_{k,q} | v_{k-1,q}).
\end{equation}
Here, the initial and transition probabilities are defined as
$p(v_{1,q}=1) = \lambda_{VR}$ and
$p(v_{k,q}=1 | v_{k-1,q}=0) = p_{01}$, respectively, 
while the remaining transition probabilities $p_{10}$, $p_{00}$, and $p_{11}$ are defined analogously and satisfy
\begin{equation}
p_{10}+p_{11}=1,
\quad
p_{00}+p_{01}=1,
\quad
\lambda_{VR} \triangleq \frac{p_{01}}{p_{01}+p_{10}},
\label{eq:transition_probs}
\end{equation}
where $\lambda_{VR}$ denotes the sparsity ratio of the VR, i.e., the proportion of RIS elements that are visible to the user.
The last equality in \eqref{eq:transition_probs} ensures that the Markov chain admits $\lambda_{VR}$ as its steady-state probability.
\begin{figure}[t]
    \centering
    \includegraphics[width=0.45\textwidth]{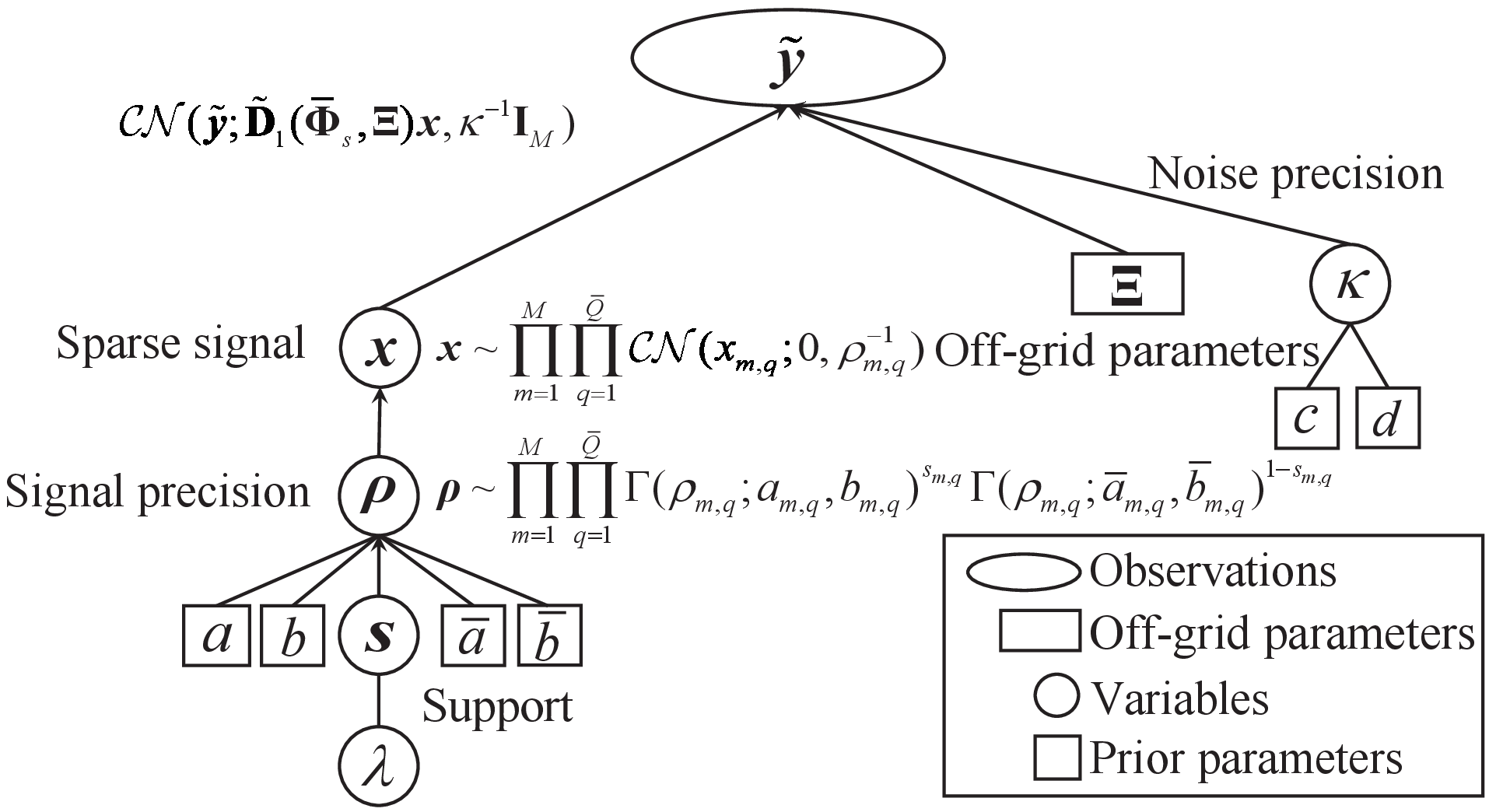}
    \caption{Joint probabilistic model of the received signal.}
    \label{fig: probability distirbution}
\end{figure}
\subsection{Joint Probabilistic Model of the Received Signal}
As shown in \eqref{tilde_y}, the received signal $\tilde{\mathbf{y}}$ can be  described by a linear observation model with additive noise.
With the probabilistic model of the cascaded channel in place, we proceed to specify the statistical characterization of the noise component, i.e., $\tilde{\mathbf{n}}$.
Specifically, the noise vector $\tilde{\mathbf{n}}$ is modeled as a complex Gaussian random vector, i.e., $\tilde{\mathbf{n}} \sim \mathcal{CN}(\mathbf{0}, \kappa^{-1}\mathbf{I})$, where $\kappa$ denotes the noise precision.
A Gamma prior is further assigned to the noise precision $\kappa$ as $p(\kappa) = \Gamma(\kappa; c, d)$,
where $c$ and $d$ are the corresponding hyperparameters.
Then, the joint distribution of all latent variables and parameters, including the received signal $\tilde{\mathbf{y}}$, the cascaded sparse channel gain vector $\boldsymbol{x}$, the subarray VR matrix $\bar{\boldsymbol{\Phi}}_s$, the hierarchical sparse variables $\bm{\rho}$ and $\boldsymbol{s}$, the off-grid parameters $\bm{\Xi}$, and the noise precision $\kappa$, can be factorized as
\begin{equation}
\begin{split}
    p(\tilde{\mathbf{y}}, \boldsymbol{x}, \bar{\boldsymbol{\Phi}}_s, \bm{\Xi}, \bm{\rho}, \boldsymbol{s}, \kappa)
    &= p(\tilde{\mathbf{y}} | \boldsymbol{x}, \bar{\boldsymbol{\Phi}}_s, \kappa, \bm{\Xi}) \\
    &\quad \times p(\boldsymbol{x}, \bm{\rho}, \boldsymbol{s}) \,
    p(\bar{\boldsymbol{\Phi}}_s)\,
    p(\kappa),
\end{split}
\label{eq:joint_distribution}
\end{equation}
where $p(\tilde{\mathbf{y}} | \boldsymbol{x}, \bar{\boldsymbol{\Phi}}_s, \kappa, \bm{\Xi})$ denotes the likelihood function.
Specifically, conditioned on $\boldsymbol{x}$, $\bar{\boldsymbol{\Phi}}_s$, $\bm{\Xi}$ and $\kappa$, the received signal $\tilde{\mathbf{y}}$ follows a complex Gaussian distribution given by
\begin{equation}
    \tilde{\mathbf{y}} \sim \mathcal{CN}(
    \tilde{\mathbf{D}}_1(\bar{\boldsymbol{\Phi}}_s, \bm{\Xi}) \boldsymbol{x},
    \kappa^{-1}\mathbf{I}
    ),
\end{equation}
where $\tilde{\mathbf{D}}_1(\bar{\boldsymbol{\Phi}}_s, \bm{\Xi})$ is constructed by row-wise concatenation of the sensing matrices corresponding to the $P$ pilot symbols. $\tilde{\mathbf{D}}_2(\boldsymbol{x}, \bm{\Xi})$ is defined in an analogous manner.
The structured sparse prior $p(\boldsymbol{x}, \bm{\rho}, \boldsymbol{s})$, and the prior of the subarray VR matrix  $p(\bar{\boldsymbol{\Phi}}_s)$ are specified in Sections~\ref{subsec: Hierarchical Sparse Probabilistic Model} and~\ref{subsec: MS}.
The resulting probabilistic dependencies among all variables are summarized in Fig.~\ref{fig: probability distirbution}.
Building upon the above probabilistic formulation, this work develops an efficient algorithm for estimating the cascaded channel parameters. To leverage both the structured sparsity of the channel and the probabilistic priors of the latent variables, a unified Bayesian inference framework is adopted. Within this framework, the TS-JBE algorithm is presented in the following section, which integrates the channel structure and prior information to jointly estimate the cascaded channel parameters.
\begin{figure}[t]
    \centering
    \includegraphics[width=0.45\textwidth]{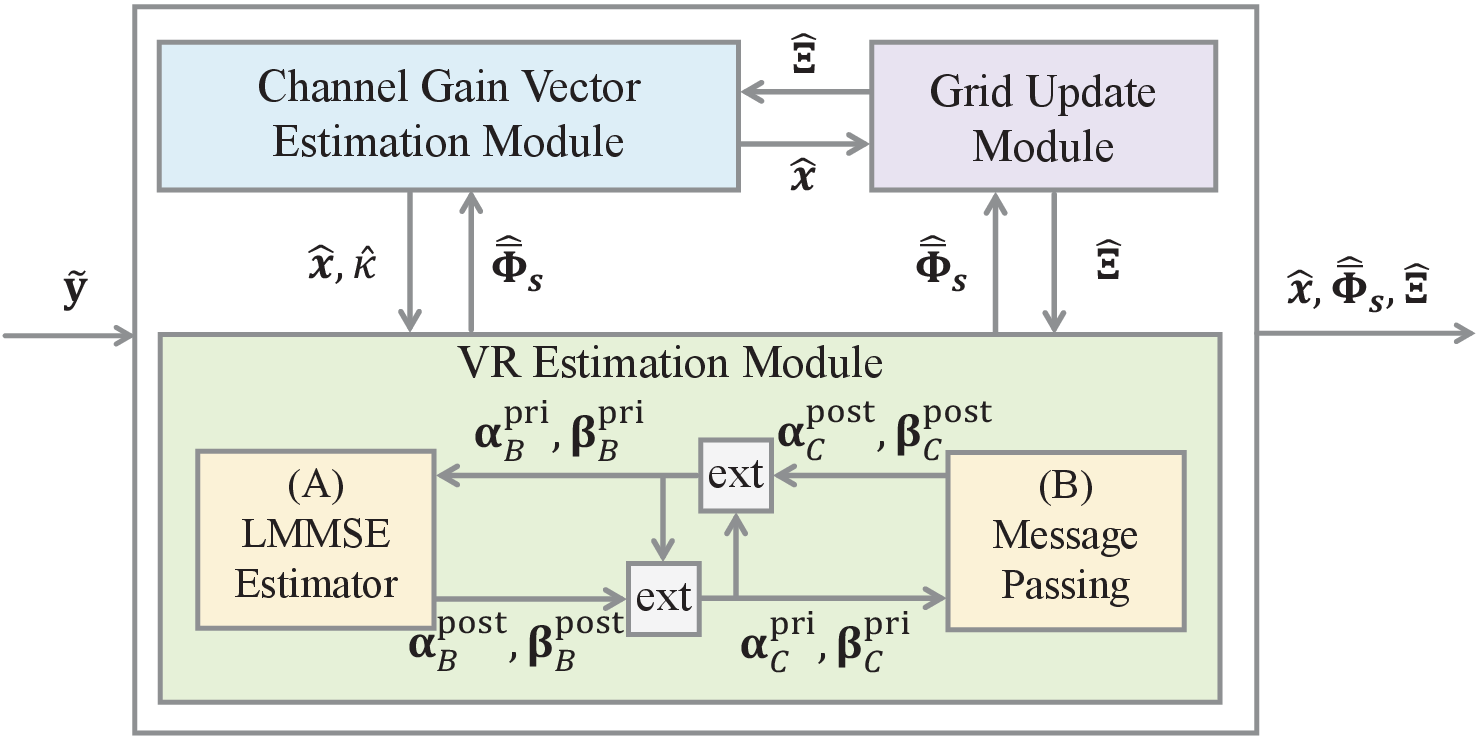}
    \caption{Flow chart of the proposed algorithm.}
    \label{fig:approach}
\end{figure}
\section{Turbo-Structured Joint Bayesian Estimation Algorithm}
\label{sec:TS-JBE}
Owing to the bilinear structure of the received signal and the disparate forms of the prior and likelihood functions, direct joint maximum a posteriori (MAP) estimation of $\boldsymbol{x}$, $\bar{\boldsymbol{\Phi}}_s$,
and $\boldsymbol{\Xi}$ is intractable.
To tackle this challenge, we develop the TS-JBE algorithm, which follows an alternating MAP inference framework to iteratively update the unknown parameters.
In particular, the joint optimization problem is formulated as
\begin{equation}
    (\boldsymbol{x}^{*}, \bar{\mathbf{\Phi}}_s^{*}, \bm{\Xi}^{*}) 
    = \arg\max_{\boldsymbol{x},\, \bar{\mathbf{\Phi}}_s, \bm{\Xi}}
    \ln p(\tilde{\mathbf{y}}, \boldsymbol{x}, \bar{\mathbf{\Phi}}_s, \bm{\Xi}, \bm{\rho}, \boldsymbol{s}, \kappa).
\end{equation}

\subsection{Outline of the TS-JBE Algorithm}
The proposed approach consists of three core modules, namely, the channel gain estimation module, the VR estimation module, and the grid update module, as illustrated in Fig.~\ref{fig:approach}. These three modules operate alternately until convergence, and their functions are outlined below.

\begin{itemize}
    \item \textbf{Channel Gain Estimation Module:}  
    Given the current VR matrix $\hat{\bar{\mathbf{\Phi}}}_s$ and the off-grid parameters $\hat{\bm{\Xi}}$, this module jointly estimates the cascaded sparse channel gain vector $\boldsymbol{x}$ and the noise precision $\kappa$ by maximizing their posterior distribution. This step leverages the structured sparsity prior of the channel as established in \eqref{eq:3LHS} and the likelihood function determined by the received signal.

    \item \textbf{VR Estimation  Module:}  
    With the estimated cascaded sparse channel gain vector $\hat{\boldsymbol{x}}$ and off-grid parameters $\hat{\bm{\Xi}}$ fixed, this module updates the subarray VR matrix $\bar{\boldsymbol{\Phi}}_s$ to identify the visible RIS elements for each path, exploiting the prior information of the visibility pattern.

    \item \textbf{Grid Update Module:}  
    To mitigate the off-grid mismatch, this module refines the off-grid parameters $\bm{\Xi}$ using a gradient descent step based on the residual error between the received signal $\tilde{\mathbf{y}}$ and its reconstruction from the current estimates of $\hat{\boldsymbol{x}}$ and $\hat{\bar{\mathbf{\Phi}}}_s$.
\end{itemize}
\vspace{-0.0cm}

\subsection{Channel Gain Vector Estimation Module}
\label{CE}

Given $\tilde{\mathbf{y}}$, 
$\hat{\bar{\mathbf{\Phi}}}_s$ and $\hat{\bm{\Xi}}$, 
the cascaded sparse channel gain vector $\boldsymbol{x}$ and the noise precision $\kappa$ 
are inferred via MAP estimation. Specifically, their estimates are given by
\begin{equation}
\begin{aligned}
\boldsymbol{x}^{*} 
&= \arg\max_{\boldsymbol{x}} 
p(\boldsymbol{x} | \tilde{\mathbf{y}}, \hat{\bar{\mathbf{\Phi}}}_s, \hat{\bm{\Xi}}), \\
\kappa^{*} 
&= \arg\max_{\kappa} 
p(\kappa | \tilde{\mathbf{y}}, \hat{\bar{\mathbf{\Phi}}}_s, \hat{\bm{\Xi}}),
\end{aligned}
\end{equation}
respectively, where $p(\boldsymbol{x}| \tilde{\mathbf{y}}, \hat{\bar{\mathbf{\Phi}}}_s, \hat{\bm{\Xi}})$ and $p(\kappa | \tilde{\mathbf{y}}, \hat{\bar{\mathbf{\Phi}}}_s, \hat{\bm{\Xi}})$ denote the posterior distributions of $\boldsymbol{x}$ and $\kappa$, 
respectively. 
According to the joint distribution in \eqref{eq:joint_distribution}, the joint estimation of $\boldsymbol{x}$ and $\kappa$ constitutes a high-dimensional Bayesian inference problem
involving multiple probabilistically coupled latent variables, including $\boldsymbol{\rho}$ and $\mathbf{s}$. To enable efficient and tractable posterior inference while exploiting the structured sparsity of $\boldsymbol{x}$, a structured-subspace-constrained
variational Bayesian inference (SSC-VBI) method is developed to infer $\boldsymbol{x}$ and $\kappa$.

Different from existing VBI-based sparse estimation methods, such as SC-VBI \cite{zhou2025SCVBI}, the proposed SSC-VBI explicitly incorporates the structured sparsity and subspace constraints of the cascaded channel into the variational inference framework. By exploiting the structural characteristics of the channel to optimize the subspace selection strategy, SSC-VBI achieves more robust and efficient posterior estimation in high-dimensional sparse scenarios, making it particularly suited for structured channel estimation.
For notational simplicity, let $\psi_j$ denote an individual variable in the set
$\bm{\Psi} \triangleq \{ \boldsymbol{x}, \bm{\rho}, \boldsymbol{s}, \kappa \}$,
and define the index set
$\mathcal{H} \triangleq \{ j | \psi_j \in \bm{\Psi} \}$.
Let $p(\bm{\Psi} | \tilde{\mathbf{y}})$ denote the posterior distribution of $\bm{\Psi}$, with $\hat{\bar{\mathbf{\Phi}}}_s$ and $\hat{\bm{\Xi}}$ treated as fixed in the obersevation matrix $
\tilde{\mathbf{D}}_1 \triangleq \tilde{\mathbf{D}}_1(\hat{\bar{\boldsymbol{\Phi}}}_s, \hat{\bm{\Xi}})
$.
According to the VBI method, the approximate posterior
$q(\bm{\Psi})$ is obtained by minimizing the KL-divergence (KLD) between
$p(\bm{\Psi} | \tilde{\mathbf{y}})$ and $q(\bm{\Psi})$
subject to a factorized form constraint, following \cite{Tzi2008VBI},
\begin{equation}
\begin{aligned}
\mathcal{A}_{\mathrm{VBI}}:\quad
\min_{q(\bm{\Psi})} \;&
\int q(\bm{\Psi})
\ln \frac{q(\bm{\Psi})}{p(\bm{\Psi} | \tilde{\mathbf{y}})}
\, d\bm{\Psi} \\
\text{s.t.}\;&
q(\bm{\Psi}) = \prod_{j \in \mathcal{H}} q(\psi_j), 
\int q(\psi_j)\, d\psi_j = 1.
\end{aligned}
\label{eq:vbi_obj}
\end{equation}
Since the KLD in $\mathcal{A}_{\mathrm{VBI}}$ is convex with respect to each individual variable $\psi_j$ while keeping all other variational distributions fixed, the variational distributions can be updated in an alternating manner to obtain a stationary solution, with the update rule given by
\begin{equation}
    q(\psi_j)
    =
    \frac{
        \exp \Big(
            \big\langle
                \ln p(\bm{\Psi}, \tilde{\mathbf{y}})
            \big\rangle_{\prod_{k \neq j} q(\psi_k)}
        \Big)
    }{
        \int 
        \exp \Big(
            \big\langle
                \ln p(\bm{\Psi}, \tilde{\mathbf{y}})
            \big\rangle_{\prod_{k \neq j} q(\psi_k)}
        \Big)
        d\psi_j
    }.
    \label{eq:vbi_general_update}
\end{equation}
Clearly, according to the above update rule, updating $q(\boldsymbol{x})$ requires computing high-dimensional matrix inverses.
To alleviate this computational burden, the SC-VBI algorithm introduces an additional structural constraint on the variational optimization problem $\mathcal{A}_{\mathrm{VBI}}$, namely,
\begin{equation}
q(\boldsymbol{x}) = \mathcal{CN}(\boldsymbol{x}; \bm{\mu}, \mathrm{diag}(\bm{\sigma}^2)),
\label{eq:add_constraint}
\end{equation}
where $\bm{\sigma}^2 = [\sigma_1^2, \sigma_2^2, \dots, \sigma_{M\bar{Q}}^2]^T$. 
This constraint enforces independence among the sparse vector elements, yielding a diagonal posterior covariance and eliminating high-dimensional matrix inversion, since the structural dependencies of $\boldsymbol{x}$ are explicitly captured by the latent variable $\boldsymbol{s}$ in the 3LHS model.
Based on the variational optimization problem~\eqref{eq:vbi_obj} with the additional constraint~\eqref{eq:add_constraint}, and the update rule~\eqref{eq:vbi_general_update}, the variational posterior updates are given as follows.
\subsubsection{Update Equation for \texorpdfstring{$\bm{x}$}{x}}
Based on Lemma~2 in \cite{zhou2025SCVBI}, the variational posterior of $\boldsymbol{x}$ takes the form
\begin{equation}
q(\boldsymbol{x}) = \mathcal{CN}(\boldsymbol{x}; \boldsymbol{\mu}, \mathrm{diag}(\bm{\sigma}^2)),
\label{eq:update x}
\end{equation}
where $
\boldsymbol{\mu} = \mathbf{G}^{-1} \tilde{\mathbf{D}}_1^{H} \tilde{\mathbf{y}}, \bm{\sigma}^2 = [ G_1^{-1},\, G_2^{-1},\, \ldots,\, G_{M\bar{Q}}^{-1} ],
$
with
$
\mathbf{G} = \mathrm{diag}(\langle \bm{\rho} \rangle) 
+ \langle \kappa \rangle \tilde{\mathbf{D}}_1^{H} \tilde{\mathbf{D}}_1.
$
\subsubsection{Update Equation for \texorpdfstring{$\bm{\rho}$}{rho}}
$q(\bm{\rho})$ can be derived as
\begin{equation}
q(\bm{\rho})
=
\prod_{m=1}^{M} \prod_{q=1}^{\bar{Q}}
\Gamma(\rho_{m,q}; \tilde{a}_{m,q}, \tilde{b}_{m,q}),
\label{eq:q_rho}
\end{equation}
where
$
\tilde{a}_{m,q}
=
\langle s_{m,q} \rangle a_{m,q}
+
\langle 1 - s_{m,q} \rangle a_{m,q}
+
1,
\tilde{b}_{m,q}
=
\langle s_{m,q} \rangle b_{m,q}
+
\langle 1 - s_{m,q} \rangle b_{m,q}
+
\langle |x_{m,q}|^2 \rangle.
$
\subsubsection{Update Equation for \texorpdfstring{$\bm{s}$}{s}} 
The posterior distribution $q(s)$
is given by
\begin{equation}
q(\boldsymbol{s})
=
\prod_{m=1}^{M} \prod_{q=1}^{\bar{Q}}
(\tilde{\lambda}_{m,q})^{s_{m,q}}
(1-\tilde{\lambda}_{m,q})^{1-s_{m,q}}.
\label{eq:q_s}
\end{equation}
\subsubsection{Update Equation for \texorpdfstring{$\kappa$}{kappa}}
The variational posterior of $\boldsymbol{\kappa}$ can be derived as
\begin{equation}
q(\kappa)
=
\Gamma(\kappa; \tilde{c}, \tilde{d}),
\label{eq:q_k}
\end{equation}
where
$
\tilde{c} = c + MP, \quad
\tilde{d}
=
d
+
\big\langle
\lVert
\tilde{\mathbf{y}} - \tilde{\mathbf{D}}_1 \boldsymbol{x}
\rVert^2
\big\rangle_{q(\boldsymbol{x})}.
$
Besides exploiting the above independence to eliminate matrix inversion in the posterior covariance $\bm{\sigma}^2$, the sparsity of $\boldsymbol{x}$ is further leveraged to identify an active index set, thereby avoiding high-dimensional matrix inversion in the posterior mean $\boldsymbol{\mu}$ update.
Specifically, based on the identified active index set $S_{\mu}$ obtained in the last iteration, the posterior
mean of $\boldsymbol{x}$ in the $i_x$-th iteration ($i_x \!\in\! \mathcal{I}_x \!\triangleq\! \{1, 2, \ldots, I_x\}$) is first updated as

\begin{equation}
\boldsymbol{u}_{i_x}^{(0)}(S_{\mu}) = \langle \kappa \rangle \mathbf{G}(S_{\mu},S_{\mu})^{-1} \tilde{\mathbf{D}}_{1}(:,S_{\mu})^H \tilde{\mathbf{y}}.
\label{eq:miu}
\end{equation}
To refine the estimation, the SC-VBI algorithm takes $\boldsymbol{u}_{i_x}^{(0)}$ as the initial solution to the following optimization problem
\begin{equation}
\min_{\boldsymbol{u}} \, \varphi(\boldsymbol{u})
\triangleq
\boldsymbol{u}^{\mathrm{H}} \mathbf{G} \boldsymbol{u}
- 2 \Re \left\{ \boldsymbol{u}^{\mathrm{H}} \tilde{\mathbf{D}}_{1}^{\mathrm{H}} \tilde{\mathbf{y}} \right\},
\label{eq:min_miu}
\end{equation}
which is equivalent to minimizing the KLD with respect to $\boldsymbol{x}$ while keeping all other variational parameters fixed, according to the update rule in~\eqref{eq:vbi_general_update}.
A gradient-based update scheme is then applied to iteratively refine $\boldsymbol{u}$, given by
\begin{equation}
\boldsymbol{u}_{i_x}^{(i_g)}
= \boldsymbol{u}_{i_x}^{(i_g-1)}
- \epsilon^{(i_g)}
\nabla_{\boldsymbol{u}} \varphi(\boldsymbol{u})
\Big|_{\boldsymbol{u} = \boldsymbol{u}_{i_x}^{(i_g-1)}},
\label{eq:grad}
\end{equation}
where $i_g \in \mathcal{I}_{g} \triangleq \{1, \ldots, I_g\}$ denotes the index of the gradient descent iteration, and $\epsilon^{(i_g)}$ is the corresponding step size. 
Accordingly, the resulting $\boldsymbol{\mu}^{(i_x)} \triangleq \boldsymbol{u}_{i_x}^{(I_g)}$ serves as the posterior mean estimate of $\boldsymbol{x}$.

In the considered problem, the sparse vector $\boldsymbol{x}$ exhibits a column-wise sparsity pattern, where each column contains at most one nonzero element. Motivated by this observation, we refine the SC-VBI by explicitly incorporating the column-wise sparsity structure of $\boldsymbol{x}$ to guide the construction of the subspace index set, thereby improving the estimation accuracy.
Specifically, let $\boldsymbol{\mu}^{(i_x-1)}$ denote the posterior mean vector obtained at the $(i_x-1)$-th iteration.
By reshaping $\boldsymbol{\mu}^{(i_x-1)}$ into matrix form as
$
\mathbf{U}^{(i_x-1)} \triangleq \mathrm{unvec}_{M \times \bar{Q}} \left(\boldsymbol{\mu}^{(i_x-1)}\right)$,
the active columns are identified according to an energy-based thresholding criterion as follows:
\begin{equation}
\mathcal{Q}_{\mu^{(i_x-1)}} \triangleq 
\left\{\, q \in \bar{\mathcal{Q}} \;\big|\; 
\big\| \mathbf{U}^{(i_x-1)}_{:,q} \big\|^2 > \varepsilon_{\mu} \,\right\},
\end{equation}
where the threshold $\varepsilon_{\mu}$ is chosen such that the selected
columns in $\mathcal{Q}_{\mu^{(i_x-1)}}$ account for a dominant portion of
the total energy of $\mathbf{U}^{(i_x-1)}$, e.g., no less than $95\%$.
For each active column $q \in \mathcal{Q}_{\mu^{(i_x-1)}}$, the dominant row index is selected as $
m_q = \arg\max_{m} | \mathbf{U}^{(i_x-1)}_{m,q} |
$. Accordingly, the final subspace index set is then defined as
\begin{equation}
S_{\mu} \triangleq 
\left\{\, (q-1)M + m_q \;\big|\; q \in \mathcal{Q}_{\mu^{(i_x-1)}} \,\right\}.
\label{eq:S_miu}
\end{equation}
The posterior mean vector $\boldsymbol{\mu}^{(i_x)}$ is then updated only over the indices in \eqref{eq:S_miu}, followed by refinement through the update rules in \eqref{eq:miu} and \eqref{eq:grad}.

\subsection{VR Estimation Module}
Given $\tilde{\mathbf{y}}$, $\hat{\boldsymbol{x}}$, $\hat{\bm{\Xi}}$ and $\hat{\kappa}$ obtained from the channel gain estimation module, the subarray VR matrix $\bar{\boldsymbol{\Phi}}_s$ is also estimated under the MAP approach according to the linear observation model 
\begin{equation}
    \tilde{\mathbf{y}}= \tilde{\mathbf{D}}_2(\hat{\boldsymbol{x}},\hat{\bm{\Xi}}) \mathrm{vec}(\bar{\boldsymbol{\Phi}}_{s}) + \tilde{\mathbf{n}} = \tilde{\mathbf{D}}_2 \mathrm{vec}(\bar{\boldsymbol{\Phi}}_{s}) + \tilde{\mathbf{n}},
\label{y_VR}
\end{equation}
where $\tilde{\mathbf{D}}_2 \triangleq \tilde{\mathbf{D}}_2(\hat{\boldsymbol{x}},\hat{\bm{\Xi}})$.
With this model, a turbo approach is employed for VR estimation, which iteratively exchanges extrinsic information between the linear Gaussian observation model and the structure-aware VR prior to achieve reliable estimation performance.
Specifically, the proposed approach consists of two interconnected submodules: one incorporates the structural prior of the VR via a message passing (MP) algorithm, while the other updates the VR posterior distribution according to the current observation model. Through iterative information exchange between these two submodules, the proposed approach reliably estimates the VR while capturing its sparsity and structural dependencies.

\subsubsection{Preprocessing and Reformulation}
Before introducing the specific estimation module, the target VR and corresponding observation model are reformulated by exploiting the sparsity of $\hat{\boldsymbol{x}}$ and the real-valued nature of the VR. 
Firstly, due to the sparsity of $\boldsymbol{x}$, only a few columns in $\bar{\boldsymbol{\Phi}}_s$ contribute to the channel. 
To reduce computational complexity, VR estimation is therefore restricted to these dominant columns.
This restriction is embedded within an alternating inference framework, in which $\boldsymbol{x}$ and $\bar{\boldsymbol{\Phi}}_s$ are alternately refined, thereby avoiding the adverse effects caused by the fixation on early-stage estimates of $\boldsymbol{x}$.
Specifically, the active column index set is determined from $\hat{\boldsymbol{x}}$ as in Section~\ref{CE}, denoted by $\mathcal{Q}_{v} \triangleq \mathcal{Q}_{\mu^{(i_x)}}$. 
Accordingly, VR estimation is performed only on the corresponding subarray VR matrix
$
\boldsymbol{\bar{\Phi}}_{s,\mathcal{Q}_{v}} = \boldsymbol{\bar{\Phi}}_s(:, \mathcal{Q}_{v}) \in \mathbb{R}^{K \times |\mathcal{Q}_{v}|},
$
with the corresponding selected measurement matrix $\tilde{\mathbf{D}}_{2,\mathcal{Q}_{v}}$.

Since $\boldsymbol{\bar{\Phi}}_{s,\mathcal{Q}{v}}$ is real-valued, the model is converted to an equivalent real-valued form $\bar{\mathbf{y}} = [\Re[\tilde{\mathbf{y}}]^T, \Im[\tilde{\mathbf{y}}]^T]^T$, with $\bar{\mathbf{D}}_{2,\mathcal{Q}v}$ and $\bar{\mathbf{n}}$ defined accordingly, yielding the real-valued mode 
\begin{equation}
    \bar{\mathbf{y}} = \bar{\mathbf{D}}_{2,\mathcal{Q}{v}} \mathrm{vec}(\boldsymbol{\bar{\Phi}}_{s,\mathcal{Q}{v}}) + \bar{\mathbf{n}}.
    \label{eq:real_model}
\end{equation}
With this reformulation, the VR estimation problem is cast into an equivalent compressed real-valued inference form, and a turbo approach consisting of two interconnected modules is developed to estimate the submatrix $\boldsymbol{\bar{\Phi}}_{s,\mathcal{Q}_{v}}$. Specifically, an LMMSE module performs linear estimation under the transformed observation model, while an MP module exploits the structural characteristics of the VR. By iteratively exchanging information between these two modules, reliable and accurate VR estimation is achieved.
\subsubsection{Module A}
In Module A, the LMMSE estimator incorporates the prior information provided by the MP module to construct a Gaussian prior for $\boldsymbol{x}$ and subsequently obtain the LMMSE estimate. 
Specifically, the prior mean and variance vectors $\boldsymbol{\alpha}_{A,pri}$ and $\boldsymbol{\beta}_{A,pri}$ are first received and used to model the submatrix $\boldsymbol{\bar{\Phi}}_{s,\mathcal{Q}_{v}}$ as a Gaussian random vector. 
Accordingly, the prior distribution of $\boldsymbol{\bar{\Phi}}_{s,\mathcal{Q}_{v}}$ is $\mathcal{N}\!\left(\mathrm{vec}(\boldsymbol{\bar{\Phi}}_{s,\mathcal{Q}_{v}});\, \boldsymbol{\alpha}_{A,pri},\, \mathrm{diag}(\boldsymbol{\beta}_{A,pri})\right).
$
At the first iteration, $\boldsymbol{\alpha}_{A,pri}$ and $\boldsymbol{\beta}_{A,pri}$ are initialized in an unbiased manner.
Given the above prior model and the transformed observation model, the posterior distribution of $\boldsymbol{\bar{\Phi}}_{s,\mathcal{Q}_{v}}$ can be obtained according to the LMMSE estimation principle. 
The resulting posterior covariance matrix and mean vector are respectively expressed as
\begin{align}
    \boldsymbol{\Gamma}_{A,post} 
    &= \left(\hat{\kappa} {\mathbf{\bar D}}_{2,\mathcal{Q}_{v}}^{\!T}\mathbf{\bar D}_{2,\mathcal{Q}_{v}} + \mathrm{diag}\!\left(\frac{1}{\boldsymbol{\beta}_{A,pri}}\right)\right)^{-1}, \label{eq:LMMSE_cov}\\
    \boldsymbol{\alpha}_{A,post} 
    &= \boldsymbol{\Gamma}_{A,post}\!\left(\frac{\boldsymbol{\alpha}_{A,pri}}{\boldsymbol{\beta}_{A,pri}} + \hat{\kappa} {\mathbf{\bar D}}_{2,\mathcal{Q}_{v}}^{\!T}\mathbf{\bar y}\right). \label{eq:LMMSE_mean}
\end{align}
Following the turbo processing principle, the posterior statistics obtained in this module cannot be directly forwarded to the subsequent message-passing module, since they are correlated with the prior information. 
To generate extrinsic information, a de-correlation operation is therefore applied~\cite{Chen2018st}. 
The resulting prior mean and variance for Module~B are given by
\begin{align}
    \boldsymbol{\alpha}_{B,pri} 
    &= \boldsymbol{\beta}_{B,pri}\!\left(\frac{\boldsymbol{\alpha}_{A,post}}{\boldsymbol{\beta}_{A,post}} - \frac{\boldsymbol{\alpha}_{A,pri}}{\boldsymbol{\beta}_{A,pri}}\right), \label{eq:decorr_alpha}\\
    \boldsymbol{\beta}_{B,pri} 
    &= \left(\frac{1}{\boldsymbol{\beta}_{A,post}} - \frac{1}{\boldsymbol{\beta}_{A,pri}}\right)^{-1}.\label{eq:decorr_beta}
\end{align}
\subsubsection{Module B}
\begin{figure}[t]
    \centering
    \includegraphics[width=0.38\textwidth]{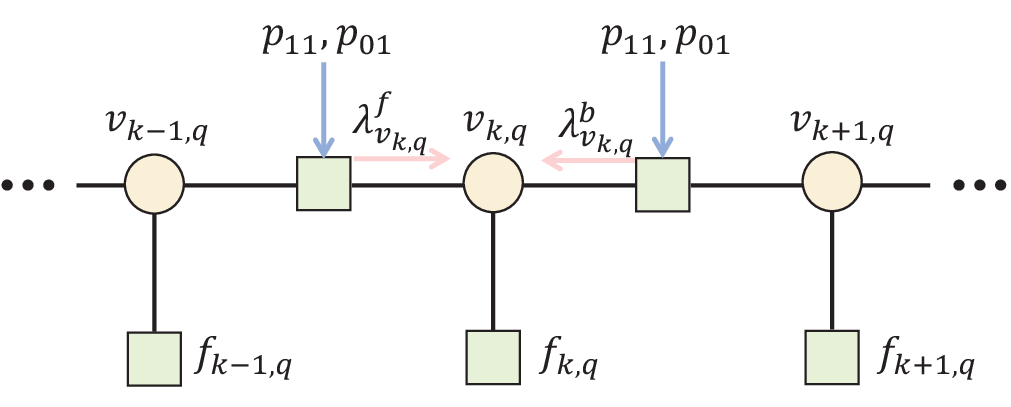}
    \caption{Factor graph of the subarray VR for the $q$-th column.}
    \label{fig:factor_graph}
\end{figure}

In Module~B, the MP estimator is built upon the LMMSE estimates by incorporating prior information on the structural characteristics of the VR, thereby achieving improved estimation performance. 
Specifically, the prior mean and variance vectors $\boldsymbol{\alpha}_{B,pri}$ and $\boldsymbol{\beta}_{B,pri}$ are interpreted as equivalent noisy observations of the binary VR variables, which are modeled as
\begin{align}
    \boldsymbol{\alpha}_{B,pri} &= \mathrm{vec}(\boldsymbol{\bar{\Phi}}_{s,\mathcal{Q}_{v}}) + \mathbf{z}, \mathbf{z} \sim \mathcal{N}\!\left(\mathbf{0},\, \mathrm{diag}(\boldsymbol{\beta}_{B,pri})\right).
\end{align}
Based on the above model, the joint distribution of $\boldsymbol{\alpha}_{B,pri}$ and $\boldsymbol{\bar{\Phi}}_{s,\mathcal{Q}_{v}}$ can be expressed as
\begin{equation}
    p(\boldsymbol{\alpha}_{B,pri}, \boldsymbol{\bar{\Phi}}_{s,\mathcal{Q}_{v}}) 
    = p(\boldsymbol{\bar{\Phi}}_{s,\mathcal{Q}_{v}}) 
      \prod_{k=1}^{K} \prod_{q=1}^{|\mathcal{Q}_{v}|}
      p(\alpha_{B,k,q}^{pri} | v_{k,q}), \label{eq:mmse_joint}
\end{equation}
where the prior distribution of $\boldsymbol{\bar{\Phi}}_{s,\mathcal{Q}_{v}}$ follows
\begin{equation}
    p(\boldsymbol{\bar{\Phi}}_{s,\mathcal{Q}_{v}}) 
    = \prod_{q=1}^{|\mathcal{Q}_{v}|} 
      p(v_{1,q}) 
      \prod_{k=2}^{K} 
      p(v_{k,q} | v_{k-1,q}). 
    \label{eq:p_Phi}
\end{equation}
According to~\eqref{eq:p_Phi}, the corresponding factor graph is illustrated in Fig.~\ref{fig:factor_graph}. 
For each factor node, the associated likelihood function is given by  
\begin{equation}
    f_{k,q} \!\triangleq\! 
    \mathcal{N}\!\left(v_{k,q}; 
    \alpha_{B,k,q}^{pri}, 
    \beta_{B,k,q}^{pri}\right)\!, k \!\in\! \{1,\dots,K\}, 
    q \!\in\! \mathcal{Q}_{v}. \label{eq:mmse_factor}
\end{equation}
Following the sum-product rule, the forward and backward messages of node $v_{k,q}$, denoted by $\lambda_{v,k,q}^{\mathrm{f}}$ and $\lambda_{v,k,q}^{\mathrm{b}}$, respectively, are updated according to~\eqref{eq:msg_forward} and~\eqref{eq:msg_backward}, as shown at the top of the next page, where the extrinsic input activation probability $\pi_{v,k,q}^{in}$ is given by
\begin{equation}
    \pi_{v,k,q}^{in} 
    = \frac{\mathcal{N}\!(1; \alpha_{B,k,q}^{pri}, \beta_{B,k,q}^{pri})}
           {\mathcal{N}\!(1; \alpha_{B,k,q}^{pri}, \beta_{B,k,q}^{pri}) 
           + \mathcal{N}\!(0; \alpha_{B,k,q}^{pri}, \beta_{B,k,q}^{pri})}. \label{eq:pi_in}
\end{equation}
By combining the forward and backward messages, the extrinsic output activation probability is obtained as
\begin{figure*}[t]
\centering
\allowdisplaybreaks
\begin{align}
    \lambda_{v,k,q}^{f} &=
    \begin{cases}
        \lambda_{VR}, & k = 1,\\[3pt]
        \dfrac{p_{01}(1 - \pi_{v,k-1,q}^{in})(1 - \lambda_{v,k-1,q}^{f}) 
        + p_{11}\pi_{v,k-1,q}^{in}\lambda_{v,k-1,q}^{f}}
        {(1 - \pi_{v,k-1,q}^{in})(1 - \lambda_{v,k-1,q}^{f}) 
        + \pi_{v,k-1,q}^{in}\lambda_{v,k-1,q}^{f}}, & k > 1,
    \end{cases} \label{eq:msg_forward}\\[4pt]
    \lambda_{v,k,q}^{b} &=
    \begin{cases}
        0.5, & k = K,\\[3pt]
        \dfrac{p_{10}(1 - \pi_{v,k+1,q}^{in})(1 - \lambda_{v,k+1,q}^{b}) 
        + p_{11}\pi_{v,k+1,q}^{in}\lambda_{v,k+1,q}^{b}}
        {(p_{00} + p_{10})(1 - \pi_{v,k+1,q}^{in})(1 - \lambda_{v,k+1,q}^{b}) 
        + (p_{11} + p_{01})\pi_{v,k+1,q}^{in}\lambda_{v,k+1,q}^{b}}, & k < K.
    \end{cases} \label{eq:msg_backward}
\end{align}
	\normalsize
	\vspace*{-8pt}
	\hrulefill
	\vspace*{-4pt}
\end{figure*}
\begin{equation}
    \pi_{v,k,q}^{out} 
    = \frac{\lambda_{v,k,q}^{f}\lambda_{v,k,q}^{b}}
           {(1 - \lambda_{v,k,q}^{f})(1 - \lambda_{v,k,q}^{b}) 
           + \lambda_{v,k,q}^{f}\lambda_{v,k,q}^{b}}. \label{eq:pi_out}
\end{equation}
At this stage, the posterior probability of $v_{k,q}=1$ is then computed as
\begin{equation}
    \hat{p}(v_{k,q}=1) 
    = \frac{\pi_{v,k,q}^{in} \pi_{v,k,q}^{out}}
           {\pi_{v,k,q}^{in} \pi_{v,k,q}^{out} 
           + (1 - \pi_{v,k,q}^{in})(1 - \pi_{v,k,q}^{out})}. \label{eq:posterior_prob}
\end{equation}
Accordingly, the posterior mean and variance of $v_{k,q}$ are computed as
\begin{align}
    \alpha_{B,k,q}^{post} 
    &= \sum_{a \in \{0,1\}} a \, \hat{p}(v_{k,q}=a), \label{eq:post_mean}\\
    \beta_{B,k,q}^{post} 
    &= \sum_{a \in \{0,1\}} (a - \alpha_{B,k,q}^{post})^2 \hat{p}(v_{k,q}=a). \label{eq:post_var}
\end{align}
Similar to Module~A, the posterior statistics are further processed to generate extrinsic information before being fed back to the LMMSE module. The corresponding extrinsic mean and variance are given by
\begin{align}
    \alpha_{A,k,q}^{pri} 
    &= \beta_{A,k,q}^{pri}
       \left(\frac{\alpha_{B,k,q}^{post}}{\beta_{B,k,q}^{post}}
       - \frac{\alpha_{B,k,q}^{pri}}{\beta_{B,k,q}^{pri}}\right), \label{eq:mmse_decor_alpha}\\
    \beta_{A,k,q}^{pri} 
    &= \left(\frac{1}{\beta_{B,k,q}^{post}}
       - \frac{1}{\beta_{B,k,q}^{pri}}\right)^{-1}. \label{eq:mmse_decor_beta}
\end{align}
Finally, a binary decision on $\boldsymbol{\bar{\Phi}}_{s,\mathcal{Q}_{v}}$ is obtained by thresholding the posterior probability at $0.5$, and the fully subarray VR matrix $\hat{\boldsymbol{\bar{\Phi}}}_{s}$ is reconstructed by setting $\hat{\boldsymbol{\bar{\Phi}}}_{s}(:,\mathcal{Q}_{v}) = \boldsymbol{\bar{\Phi}}_{s,\mathcal{Q}_{v}}$.
\subsection{Grid Update Module}
Given the estimated VR $\hat{\bar{\boldsymbol{\Phi}}}_s$ and the cascaded channel gain $\hat{\boldsymbol{x}}$, the off-grid parameters $\boldsymbol{\Xi}$ are refined by solving the following 
maximum-likelihood (ML) problem:
\begin{equation}
\begin{aligned}
\max_{\Delta {\bm{\bar{\varphi}}}, \Delta {\mathbf{\bar{r}}}, \bm{\vartheta}} 
L(\Delta \bm{\bar{\varphi}}, \Delta {\mathbf{\bar{r}}}, \Delta \bm{\vartheta}) 
&= - \sum_{p=1}^{P} \Big\| \mathbf{y}_p - (\boldsymbol{\eta}_p^T \otimes \mathbf{I}_M) \\
&\quad \times ((\mathbf{Q}(\Delta \bm{\bar{\varphi}}, \Delta {\mathbf{\bar{r}}}) 
\odot \hat{\bar{\boldsymbol{\Phi}}}) \\
&\quad \otimes \mathbf{F}_M(\Delta \bm{\vartheta})) \, \hat{\boldsymbol{x}} 
\Big\|^2,
\end{aligned}
\label{eq:grid_obj}
\end{equation}
where $\hat{\bar{\boldsymbol{\Phi}}} \triangleq \hat{\bar{\boldsymbol{\Phi}}}_{s} \otimes \mathbf{1}_{\frac{N}{K} \times 1}$.
For convenience, define 
\(\mathbf{F}_p \triangleq \boldsymbol{\eta}_p^{T} \otimes \mathbf{I}_M\), 
\(\bar{\mathbf{F}}_M \triangleq \mathbf{F}_M(\Delta\boldsymbol{\vartheta})\), 
and 
\(\bar{\mathbf{Q}}_{V} \triangleq \mathbf{Q}(\Delta\bar{\boldsymbol{\varphi}},\, \Delta\bar{\mathbf{r}}) 
\odot \hat{\bar{\boldsymbol{\Phi}}}\).
Then, the objective function of problem \eqref{eq:grid_obj} can be rewritten as
\begin{equation}
\begin{aligned}
L\bigl(\Delta\bar{\boldsymbol{\varphi}}, \Delta\bar{\mathbf{r}},\Delta\boldsymbol{\vartheta}\bigr)
\!=\! - \sum_{p=1}^{P} \Big\| 
\mathbf{y}_p 
- \mathbf{F}_p
(\bar{\mathbf{Q}}_{V} \otimes \bar{\mathbf{F}}_M) \hat{\boldsymbol{x}}
\Big\|^{2}.
\end{aligned}
\label{eq:grid_expand}
\end{equation}
To facilitate computing the derivatives of $L\bigl(\Delta\bar{\boldsymbol{\varphi}}, \Delta\bar{\mathbf{r}},\Delta\boldsymbol{\vartheta}\bigr)$ with respect to the off-grid set $\boldsymbol{\Xi}$, 
we introduce intermediate terms.
For the $q$-th near-field angular component, define
\begin{equation}
\begin{aligned}
\boldsymbol{y}^{(\Delta\bar{\varphi}_q)}_p 
&\!=\! \bigl( (\mathbf{a}_q^{(\bar{\varphi}_q)} \odot \hat{\bar{\boldsymbol{\phi}}}_q) 
\otimes \bar{\mathbf{F}}_M \bigr)^{H} \mathbf{F}_p^{H} \mathbf{y}, \\
\boldsymbol{z}^{(\Delta\bar{\varphi}_q)}_p 
&\!=\! \bigl( (\mathbf{a}_q^{(\bar{\varphi}_q)} \odot \hat{\bar{\boldsymbol{\phi}}}_q) 
\otimes \bar{\mathbf{F}}_M \bigr)^{H} \mathbf{F}_p^{H} \mathbf{F}_p 
(\bar{\mathbf{Q}}_{V} \otimes \bar{\mathbf{F}}_M) \hat{\boldsymbol{x}}.
\end{aligned}
\end{equation}
Let the derivative of the ARV $\mathbf{a}_q$ with respect to the off-grid
angular perturbation $\Delta\bar{\varphi}_q$ be $\mathbf{a}_q^{(\Delta\bar{\varphi}_q)}
=
\frac{\partial \mathbf{a}([\bar{\varphi}_q+\Delta\bar{\varphi}_q,\bar{r}_q])}
{\partial \Delta\bar{\varphi}_q}.
$
Using the above intermediate terms, the gradient of
$L(\Delta\bar{\boldsymbol{\varphi}},\Delta\bar{\mathbf{r}},\Delta\boldsymbol{\vartheta})$
with respect to $\Delta\bar{\varphi}_q$ can be compactly written as
\begin{equation}
\frac{\partial L}{\partial \Delta \bar{\varphi}_q}
=
2\,\Re \sum_{p=1}^{P}
\hat{\boldsymbol{x}}_{(q-1)M+1:qM}^{H}
\big(
\boldsymbol{y}^{(\Delta\bar{\varphi}_q)}_p
-
\boldsymbol{z}^{(\Delta\bar{\varphi}_q)}_p
\big).
\end{equation}
Similarly, the gradients with respect to
$\Delta\bar{r}_q$ and $\Delta\vartheta_m$
can be derived respectively.
Finally, the off-grid parameters are sequentially refined through gradient-based updates as follows:
\begin{align}
\Delta \bm{\bar{\varphi}}^{(i+1)} 
&= \Delta \bm{\bar{\varphi}}^{(i)} 
- \varepsilon_{\bar{\varphi}} \,
\frac{\partial L(\Delta \bm{\bar{\varphi}}^{(i)}, \Delta \mathbf{\bar{r}}^{(i)}, \Delta \bm{\vartheta}^{(i)})}
{\partial \Delta \bm{\bar{\varphi}}^{(i)}}, \label{eq:update_omega} \\
\Delta \mathbf{\bar{r}}^{(i+1)} 
&= \Delta \mathbf{\bar{r}}^{(i)} 
- \varepsilon_{\bar{r}} \,
\frac{\partial L(\Delta \bm{\bar{\varphi}}^{(i+1)}, \Delta \mathbf{\bar{r}}^{(i)}, \Delta \bm{\vartheta}^{(i)})}
{\partial \Delta \mathbf{\bar{r}}^{(i)}}, \label{eq:update_r} \\
\Delta \bm{\vartheta}^{(i+1)} 
&= \Delta \bm{\vartheta}^{(i)} 
- \varepsilon_{\vartheta} \,
\frac{\partial L(\Delta \bm{\bar{\varphi}}^{(i+1)}, \Delta \mathbf{\bar{r}}^{(i+1)}, \Delta \bm{\vartheta}^{(i)})}
{\partial \Delta \bm{\vartheta}^{(i)}}, \label{eq:update_phi}
\end{align}
where the step sizes $\varepsilon_{\bar{\varphi}}$, $\varepsilon_{\bar{r}}$, and $\varepsilon_{\vartheta}$ 
are adaptively selected via a backtracking line search \cite{nocedal2006numerical}.

\renewcommand\baselinestretch{\linspreadalgr}\selectfont 
\begin{algorithm}[htb]
	\caption{Proposed TS-JBE algorithm}
	\label{JE algorithm}
	\textbf{Input}: receive signal $\tilde{\mathbf{y}}$, outer iterations $I_{out}$, channel vector iteration number $I_x$, VR iteration number $I_v$, gradient update iteration number $I_g$. \\
    \textbf{Output}: channel sparse vector $\hat{\boldsymbol{x}}$, VR matrix $\hat{\bar{\mathbf{\Phi}}}_s$, off-grid parameters $\hat{\boldsymbol{\Xi}}$.
	\begin{algorithmic}[1]
		\STATE { \textbf{For} $i_{out} = 1, \dots, I_{out}$ }
        \STATE {\;\; Initialize $q(\boldsymbol{s})$, $q(\boldsymbol{\rho})$,$q(\boldsymbol{\kappa})$. }\!\!\!
        \STATE {\textbf{Channel Gain Estimation Module}}
        \STATE {\;\; \textbf{For} $i_{x} = 1, \dots, I_{x}$}
        \STATE {\;\;\;\; Update $q(\boldsymbol{x})$ using \eqref{eq:update x}, where the posterior mean 
        $\boldsymbol{\mu}$ \\\;\;\;\;\;is obtained by performing the gradient update \eqref{eq:grad} \\\;\;\;\;
        with the initial point from \eqref{eq:miu} and \eqref{eq:S_miu} for $I_{g}$ times.}
        \STATE {\;\;\;\; Update $q(\boldsymbol{\rho})$ using \eqref{eq:q_rho}.}
        \STATE {\;\;\;\; Update $q(\boldsymbol{s})$ using \eqref{eq:q_s}.}
        \STATE {\;\;\;\; Update $q(\boldsymbol{\kappa})$ using \eqref{eq:q_k}.}
		\STATE {\;\; \textbf{end For} }
        \STATE {\;\; \text{Estimate} $\hat{\boldsymbol{x}} = \boldsymbol{\mu}^{(I_{x})},\hat{\boldsymbol{\kappa}}=\frac{\tilde{c}}{\tilde{d}} $.}
        \STATE {\;\; Transform receive signal model as \eqref{eq:real_model}.}
        \STATE {\textbf{VR Estimation Module}}
        \STATE {\;\; \textbf{For} $i_{v} = 1, \dots, I_{v}$}
        \STATE {\;\;\;\; \textbf{\%Module A: LMMSE}}  
        \STATE {\;\;\;\; \text{Compute} $\boldsymbol{\Gamma}_A^{post}$, $\boldsymbol{\alpha}_A^{post}$ using \eqref{eq:LMMSE_cov} and \eqref{eq:LMMSE_mean}.}
        \STATE {\;\;\;\; \text{Compute} $\boldsymbol{\alpha}_B^{pri}$, $\boldsymbol{\beta}_B^{pri}$ using \eqref{eq:decorr_alpha} and \eqref{eq:decorr_beta}.}
        \STATE {\;\;\;\; \textbf{\%Module B: MP}}  
        \STATE {\;\;\;\; \text{Compute} $\pi_{\nu, k, q}^{in}$ using \eqref{eq:pi_in}.}
        \STATE {\;\;\;\; \text{Compute} $\lambda_{\nu, k, q}^f$, $\lambda_{\nu, k, q}^b$ using \eqref{eq:msg_forward} and \eqref{eq:msg_backward}.}
        \STATE {\;\;\;\; \text{Compute} $\pi_{\nu, k, q}^{out}$ using \eqref{eq:pi_out}.}
        \STATE {\;\;\;\; \text{Compute} $\boldsymbol{\alpha}_B^{post}$, $\boldsymbol{\beta}_B^{post}$ using \eqref{eq:post_mean} and \eqref{eq:post_var}.}
        \STATE {\;\;\;\; \text{Compute} $\boldsymbol{\alpha}_A^{pri}$, $\boldsymbol{\beta}_A^{pri}$ using \eqref{eq:mmse_decor_alpha} and \eqref{eq:mmse_decor_beta}.}
        \STATE {\;\;\;\; \text{Estimate} $\hat{\bar{\mathbf{\Phi}}}_s$ using a 0.5 threshold.}
        \STATE {\;\; \textbf{end For} }
        \STATE {\textbf{Grid Update Module}}
        \STATE {\;\; \text{estimate} $\bar{\boldsymbol{\Xi}}$ using  \eqref{eq:update_omega},\! \eqref{eq:update_r} and \eqref{eq:update_phi}.}
		\STATE {\textbf{end For} }
	\end{algorithmic}
\end{algorithm}
\vspace{-0.1cm}
\renewcommand\baselinestretch{\linspread}\selectfont
\subsection{Complexity Analysis}

We summarize the proposed TS-JBE approach in Algorithm~1. In the channel gain vector estimation module, SSC-VBI avoids large-scale matrix inversion by restricting the operation to a reduced subspace of size $\mathbb{C}^{|S_{\mu}| \times |S_{\mu}|}$, combined with several matrix–vector multiplications, achieving a per-iteration complexity of $\mathcal{O}(M^{2}\bar{Q} + |S_{\mu}|^{3})$. In the VR estimation module, LMMSE estimator requires a matrix inversion of size $\mathbb{C}^{(K|\mathcal{Q}_{v}|) \times (K|\mathcal{Q}_{v}|)}$, resulting in a complexity of $\mathcal{O}((K|\mathcal{Q}_{v}|)^{3})$, whereas the MP update exhibits linear complexity. In the grid update module, the dominant cost arises from computing gradients with respect to the polar grid parameters $[\bar{\boldsymbol{\varphi}}, \bar{\boldsymbol{r}}]$, resulting in a complexity of $\mathcal{O}(\bar{Q}^{2})$. Denoting the iteration numbers of the channel gain and VR modules by $I_{x}$ and $I_{v}$, respectively, the overall per-outer-iteration complexity of TS-JBE is $\mathcal{O}\bigl(I_{x}(M^{2}\bar{Q} + |S_{\mu}|^{3}) + I_{v}(K|\mathcal{Q}_{v}|)^{3} + \bar{Q}^{2}\bigr)$.

\section{Simulation results}
\label{sec:simulation}
In this section, numerical results are provided to validate the effectiveness of the proposed TS-JBE algorithm. We consider a RIS-assisted MIMO system operating at a carrier frequency of $f=28$ GHz under the TDD mode.
The BS is equipped with a ULA of $M=16$ antennas with inter-element spacing $d=\frac{\lambda}{2}$. The RIS comprises $N=128$ reflecting elements arranged as a ULA with the same spacing, and a single-antenna user is assumed.
The BS and RIS are aligned along the $y$- and $x$-axes with reference positions $[-90\mathrm{m},-30\mathrm{m}]^{T}$ and $[0\mathrm{m},0\mathrm{m}]^{T}$, respectively.
The user’s distance to the RIS is between \(10~\mathrm{m}\) and 
\(20~\mathrm{m}\).
Based on the array aperture and the carrier wavelength, the Rayleigh distance of
the RIS is approximately \(87.5~\mathrm{m}\).
The distance from the BS to the RIS is about \(95.2~\mathrm{m}\), indicating that
the BS lies in the far-field of the RIS, whereas the user is located in
the near-field region.
This configuration is consistent with the considered hybrid-field propagation scenario.
The large-scale path loss follows
$PL=\alpha_{PL}+10\beta_{PL}\log_{10}(d)+\xi_{PL}$~(dB), $\xi_{PL}\sim\mathcal{N}(0,\sigma_{PL}^{2})$,
with the model and parameters adopted from~\cite{A2014loss}.
Unless stated otherwise, $L_{RB}\!=\!L_{U}\!=\!3$ paths are considered, the polar-domain dictionary has $\bar{Q}\!=\!256$ points, the RIS is split into $K\!=\!8$ subarrays, the VR sparsity ratio is $0.875$, and SNR is $20$~dB.
In this paper, the NMSE of the cascaded channel is adopted as the performance metric. The cascaded channel vector is defined as $\mathbf{h}_c = \mathrm{vec}(\mathbf{H} \mathrm{diag}(\mathbf{h}_u)) $. 
Accordingly, the NMSE is defined as $
\| \hat{\mathbf{h}}_{c} - \mathbf{h}_{c} \|^2 \big/
\| \mathbf{h}_{c} \|^2,
$
where $\hat{\mathbf{h}}_{c} = ( ( \mathbf{Q}(\Delta \bar{\boldsymbol{\varphi}}, \Delta \bar{\mathbf{r}}) 
\odot \hat{\bar{\boldsymbol{\Phi}}} )
\otimes \mathbf{F}_{M}(\Delta \boldsymbol{\vartheta}) )
\hat{\boldsymbol{x}}$ denotes the reconstructed cascaded channel vector. 
For performance comparison, we consider the following baseline algorithms:
\begin{itemize}
    \item \textbf{OMP}: This method assumes that all RIS elements are visible to the signal and applies OMP to estimate the cascaded channel.

    \item \textbf{SBL}: Similar to OMP, but uses sparse Bayesian learning (SBL) for cascaded channel estimation.

    \item \textbf{RFSBL}: The robust fast SBL (RFSBL) algorithm estimates the cascaded sparse channel gains, VRs, and off-grid parameters through a two-stage procedure~\cite{Yu2023nf}.

    \item \textbf{TS-JBE without VR estimation}: This benchmark employs the proposed TS-JBE algorithm while disabling the VR estimation module. Only the cascaded sparse channel gains and the off-grid parameters are estimated.

    \item \textbf{Oracle LS}: LS estimation with perfect knowledge of path distances, angles, scatterer locations, and VRs, serving as a performance lower bound.

\end{itemize}

\begin{figure}[t]
    \centering
    \includegraphics[width=0.28\textwidth]{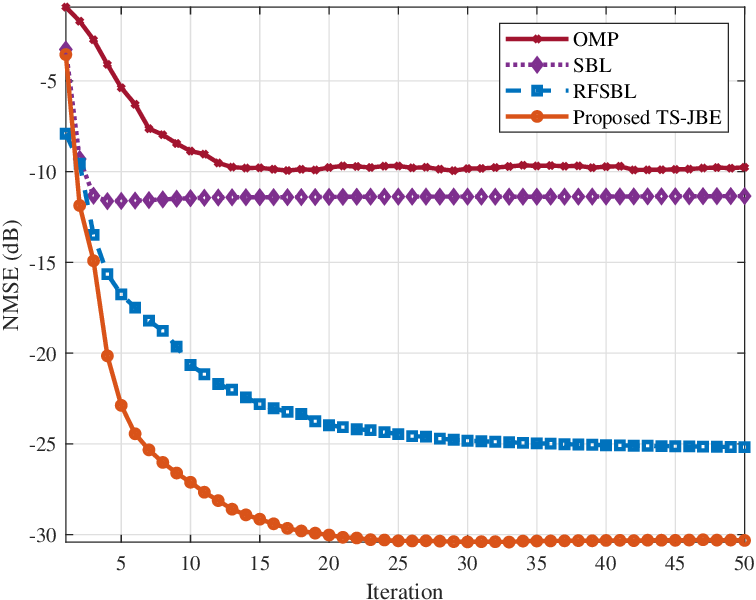}
    \caption{Convergence behavior of the considered algorithms}
    \label{fig:Convergence performance}
\end{figure}
\begin{figure}[t]
    \centering
    \includegraphics[width=0.28\textwidth]{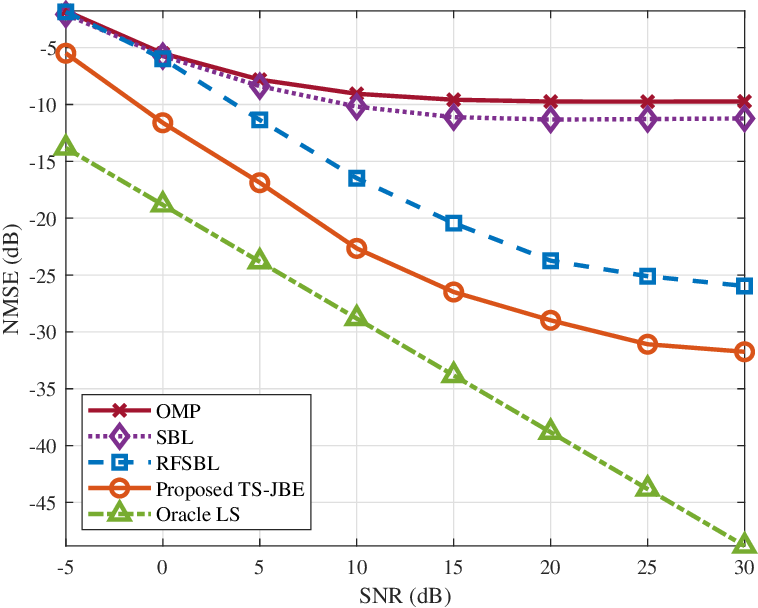}
    \caption{NMSE performance of the considered algorithms
versus SNR.}
    \label{fig:NMSE-dB}
\end{figure}
Before presenting the performance results, Fig. \ref{fig:Convergence performance} verifies the convergence of the considered algorithms.
The proposed algorithm shows a rapid NMSE reduction in early iterations, converging to a stable value after around 25 iterations, with a monotonically decreasing trend indicating good stability. This favorable convergence behavior stems from the proposed joint iterative approach, in which the cascaded sparse channel gains, VRs, and off-grid parameters are alternately updated, enabling consistent information exchange and progressive refinement across iterations.
In comparison, SBL and OMP converge faster due to simpler iterative structures without joint VR and off-grid estimation, while RFSBL, also using multi-parameter alternating updates, converges more slowly, stabilizing after about 30 iterations.

Fig. \ref{fig:NMSE-dB} compares the NMSE performance of different algorithms under varying SNRs. As can be observed, estimation accuracy improves for all schemes as SNR increases. The proposed TS-JBE consistently outperforms OMP, SBL, and RFSBL across the entire SNR range. OMP and SBL do not consider non-stationarity or off-grid effects, causing performance to saturate in the medium-SNR regime, with little improvement beyond 15 dB. Both RFSBL and TS-JBE explicitly model the VR and off-grid effects, yielding much better performance than OMP and SBL. However, RFSBL's two-stage estimation is prone to error propagation, limiting its final accuracy. In contrast, TS-JBE employs joint estimation with a three-layer prior and Markov-structured constraints, providing stable gains and accurate characterization of the channel and VR structures. Compared with the Oracle LS benchmark, TS-JBE maintains an approximately constant gap in low-to-medium SNRs, demonstrating robustness to noise and stable estimation behavior.

\begin{figure}[t]
    \centering
    \includegraphics[width=0.28\textwidth]{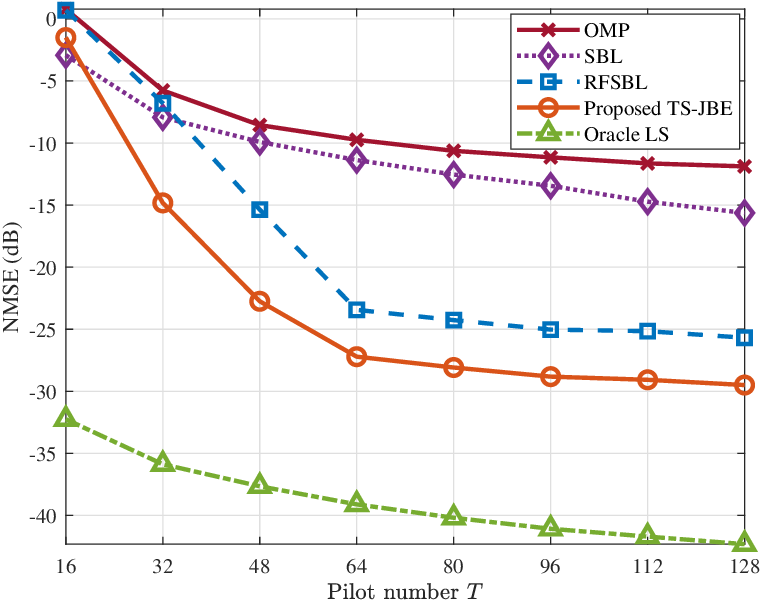}
    \caption{NMSE performance under different pilot numbers.}
    \label{fig:NMSE-pilot}
\end{figure}
\begin{figure}[t]
    \centering
    \includegraphics[width=0.28\textwidth]{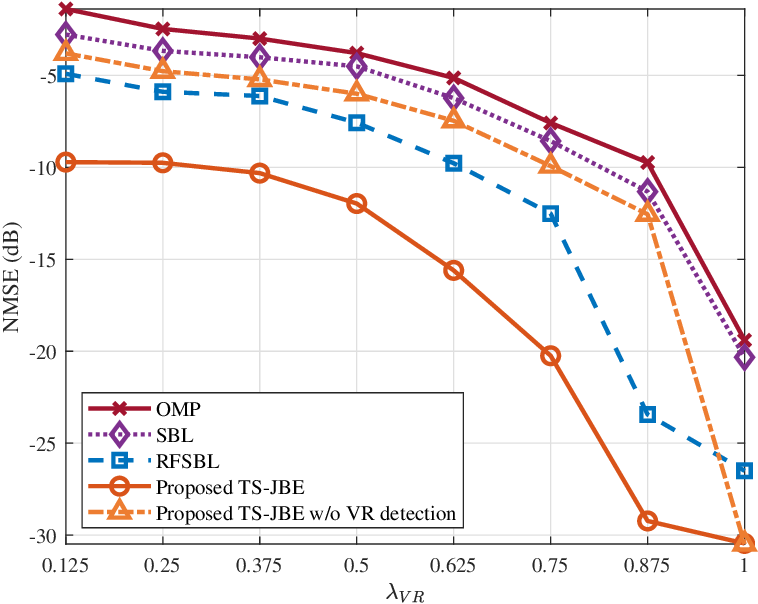}
    \caption{NMSE performance under different VR sparsity ratios.}
    \label{fig:NMSE-VR}
\end{figure}
Fig. \ref{fig:NMSE-pilot} shows the NMSE performance of different algorithms versus the pilot number $T$. The NMSE of all schemes decreases monotonically as $T$ increases. The proposed TS-JBE consistently outperforms all benchmarks when $T>20$, maintaining an advantage over RFSBL across the entire range. Beyond $T=64$, its performance saturates, indicating that satisfactory estimation accuracy is achieved with 64 pilots. In contrast, OMP and SBL reduce NMSE much more slowly and perform significantly worse, as they do not account for channel non-stationarity and off-grid effects and thus cannot fully exploit longer pilot sequences.

Fig. \ref{fig:NMSE-VR} shows the NMSE performance of different algorithms under varying VR sparsity ratios $\lambda_{VR}$. As $\lambda_{VR}$ increases, the NMSE of all schemes generally decreases due to richer observations from more visible RIS elements. Algorithms explicitly modeling the VR consistently outperform baselines assuming a fully visible RIS, such as OMP and SBL, highlighting the benefit of VR modeling. Among these, the proposed TS-JBE consistently maintains performance gains over RFSBL even under low visibility conditions, indicating the stability of the algorithm.
Furthermore, incorporating VR estimation significantly improves performance in partially visible scenarios, whereas this gain disappears when $\lambda_{VR}=1$, where all RIS elements are visible and explicit VR modeling is unnecessary.

\section{Conclusion}
\label{sec:conclusion}
This paper investigated non-stationary cascaded channel estimation in RIS-assisted millimeter-wave systems under a hybrid-field propagation environment. Exploiting the cascaded channel structure, a compression strategy based on a visibility-matrix-weighted low-dimensional polar-domain dictionary was designed to construct a reduced-dimensional sparse bilinear representation. 
Based on this representation, a TS-JBE approach was proposed to jointly estimate the cascaded sparse channel gains, VRs, and off-grid parameters. Simulations showed that the proposed approach significantly improves NMSE compared with existing methods.

\section*{Appendix}
\subsection{Proof of Proposition 1}
The $\big(n_x,(n_{\bar{y}}-1)N+n_y\big)$-th element of
$\bar{\mathbf{W}}\ast\mathbf{F}_N^{\ast}$, given by
$\bar{\mathbf{W}}(n_x,n_{\bar{y}})\,\mathbf{F}_N^{\ast}(n_x,n_y)$, is expressed as
\begin{equation}
\boldsymbol{\phi}_{n_x,n_{\bar{y}}}
e^{
-j\frac{2\pi}{\lambda}
\Big(
-(n_x-1)d\big(\vartheta_{n_{\bar{y}}}-\varphi_{n_y}\big)
+ \frac{d^2 (n_x-1)^2}{2 r_{n_{\bar{y}}}}\big(1-\vartheta_{n_{\bar{y}}}^2\big)
\Big)},
\label{eq:WF}
\end{equation}
where $n_x,n_y\in\mathcal{N}$ and $n_{\bar{y}}\in\bar{\mathcal{N}}$.
Define $\bar{\varphi} = \vartheta_{n_{\bar{y}}} - \varphi_{n_y}$ and
$\bar{r} = \frac{r_{n_{\bar{y}}}(1 - \bar{\varphi}^2)}{1 - \vartheta_{n_{\bar{y}}}^2}$.
Then, \eqref{eq:WF} can be rewritten as
\begin{equation}
\boldsymbol{\phi}_{n_x,n_{\bar{y}}}  
e^{-j \frac{2 \pi}{\lambda} \Big(-(n_x -1) d \, \bar{\varphi} + \frac{d^2(n_x -1)^2}{2 \bar{r}} (1-\bar{\varphi}^2) \Big)}.
\label{eq:transform_WF}
\end{equation}
The phase term in \eqref{eq:transform_WF} matches the analytical form of a near-field ARV, while the amplitude term corresponds to the VR indicator.
Consequently, each column of $\bar{\mathbf{W}} \ast \mathbf{F}_N^\ast$ can be interpreted as a near-field ARV parameterized by $(\bar{\varphi}, \bar{r})$ and weighted by the associated VR indicator.
By constructing a new polar-domain grid
$\{(\bar{\varphi}_q, \bar{r}_q)\}_{q=1}^{\bar{Q}}$ over $(\bar{\varphi}, \bar{r})$,
where $\bar{Q}$ denotes the number of grid points,
the associated ARVs form an equivalent near-field dictionary $\mathbf{Q}$.
Meanwhile, a VR matrix $\bar{\boldsymbol{\Phi}}$ is introduced to maintain a one-to-one correspondence after the dictionary transformation.
As a result, $\bar{\mathbf{W}} \ast \mathbf{F}_N^\ast$ admits the equivalent representation $\mathbf{Q} \odot \bar{\boldsymbol{\Phi}}$.
%%%%%%%%%%%%%%%%%%%%%%%%%%%%%% Bibliography %%%%%%%%%%%%%%%%%%%%%%%%%%%%%%%%%%%%
\bibliographystyle{IEEEtran}
%\bibliography{IEEEabrv,bib}
\bibliography{bib_channel_estimation}

@ARTICLE{cui2022nearfield,
	author={Cui, Mingyao and Dai, Linglong},
	journal={IEEE Trans. Commun.}, 
	title={Channel Estimation for Extremely Large-Scale {MIMO}: Far-Field or Near-Field?}, 
	year={2022},
	volume={70},
	number={4},
	pages={2663-2677},
	  month={Apr.},
	doi={10.1109/TCOMM.2022.3146400}}

@ARTICLE{Selvan2017fraun,
	author={Selvan, Krishnasamy T. and Janaswamy, Ramakrishna},
	journal={IEEE Antennas Propag. Mag.}, 
	title={Fraunhofer and {Fresnel} Distances: Unified derivation for aperture antennas}, 
	year={2017},
	volume={59},
	number={4},
	pages={12-15},
	doi={10.1109/MAP.2017.2706648},
	ISSN={1558-4143},
	month={Aug.},}

@ARTICLE{xu2023successive,
  author={Xu, Wenkang and Liu, An and Zhou, Bingpeng and Zhao, Min-Jian},
  journal={IEEE Trans. Wireless Commun.}, 
  title={Successive Linear Approximation {VBI} for Joint Sparse Signal Recovery and Dynamic Grid Parameters Estimation}, 
  year={2025},
  volume={24},
  number={11},
  pages={9645-9659},
  month={Nov.},
  doi={10.1109/TWC.2025.3574489}}

@ARTICLE{Han2022Localization,
  author={Han, Yu and Jin, Shi and Wen, Chao-Kai and Quek, Tony Q. S.},
  journal={IEEE J. Sel. Top. Signal Process.}, 
  title= {Localization and Channel Reconstruction for Extra-Large {RIS}-Assisted Massive {MIMO} Systems},

  year={2022},
  volume={16},
  number={5},
  pages={1011-1025},
  month={Aug.},
  doi={10.1109/JSTSP.2022.3174654}}

@ARTICLE{Yu2023nf,
  author={Yu, Xiao and Shen, Wenqian and Zhang, Rui and Xing, Chengwen and Quek, Tony Q. S.},
  journal={IEEE Trans. Commun.}, 
  title={Channel Estimation for {XL-RIS}-Aided Millimeter-Wave Systems}, 
  year={2023},
  volume={71},
  number={9},
  pages={5519-5533},
  month={Sep.},
  doi={10.1109/TCOMM.2023.3286450}}

@ARTICLE{zhou2025SCVBI,
  author={Liu, An and Zhou, Yufan and Xu, Wenkang},
  journal={IEEE Trans. Signal Process.}, 
  title={Subspace Constrained Variational {Bayesian} Inference for Structured Compressive Sensing With a Dynamic Grid}, 
  year={2025},
  volume={73},
  number={},
  month={Jan.},
  pages={781-794},
  doi={10.1109/TSP.2025.3532953}}

@ARTICLE{Chen2018st,
  author={Chen, Lei and Liu, An and Yuan, Xiaojun},
  journal={IEEE Trans. Veh. Technol.	}, 
  title={Structured Turbo Compressed Sensing for Massive {MIMO} Channel Estimation Using a {Markov} Prior}, 
  year={2018},
  volume={67},
  number={5},
  month={May},
  pages={4635-4639},
  doi={10.1109/TVT.2017.2787708}}

@ARTICLE{A2014loss,
  author={Akdeniz, Mustafa Riza and Liu, Yuanpeng and Samimi, Mathew K. and Sun, Shu and Rangan, Sundeep and Rappaport, Theodore S. and Erkip, Elza},
  journal={IEEE J. Sel. Areas Commun.}, 
  title={Millimeter Wave Channel Modeling and Cellular Capacity Evaluation}, 
  year={2014},
  volume={32},
  number={6},
  month={Jun.},
  pages={1164-1179},
  doi={10.1109/JSAC.2014.2328154}}

@ARTICLE{liu2020HLS,
  author={Liu, An and Liu, Guanying and Lian, Lixiang and Lau, Vincent K. N. and Zhao, Min-Jian},
  journal={IEEE Trans. Wireless Commun.	}, 
  title={Robust Recovery of Structured Sparse Signals With Uncertain Sensing Matrix: A Turbo-{VBI} Approach}, 
  year={2020},
  volume={19},
  number={5},
  month={May},
  pages={3185-3198},
  doi={10.1109/TWC.2020.2971193}}

@ARTICLE{Lu2024NF,
  author={Lu, Haiquan and Zeng, Yong and You, Changsheng and Han, Yu and Zhang, Jiayi and Wang, Zhe and Dong, Zhenjun and Jin, Shi and Wang, Cheng-Xiang and Jiang, Tao and You, Xiaohu and Zhang, Rui},
  journal={IEEE Commun. Surveys Tuts.}, 
  title={A Tutorial on Near-Field {XL-MIMO} Communications Toward 6{G}}, 
  year={2024},
  volume={26},
  number={4},
  month ={4th Quart.},
  pages={2213-2257},
  doi={10.1109/COMST.2024.3387749}}

@INPROCEEDINGS{Zhou2024VR,
  author={Zhou, Chao and You, Changsheng and Gong, Shiqi and Lyu, Bin and Zheng, Beixiong and Gong, Yi},
  booktitle={Proc. IEEE 16th Int. Conf. Wireless Commun. Signal Process. (WCSP) }, 
  title={Channel Estimation for {XL-IRS} Assisted Wireless Systems with Double-sided Visibility Regions}, 
  year={2024},
  volume={},
  number={},
  month ={Oct.},
  pages={456-461},
  doi={10.1109/WCSP62071.2024.10827219}}

@book{zhang2017matrix,
  title={Matrix analysis and applications},
  author={Zhang, Xianda},
  year={2017},
  publisher={Cambridge Univ. Press},
  address   = {Cambridge, U.K.}
}

@ARTICLE{Tang2024VR,
  author={Tang, Anzheng and Wang, Jun-Bo and Pan, Yijin and Zhang, Wence and Zhang, Xiaodan and Chen, Yijian and Yu, Hongkang and de Lamare, Rodrigo C.},
  journal={IEEE Trans. Commun.}, 
  title={Joint Visibility Region and Channel Estimation for Extremely Large-Scale {MIMO} Systems}, 
  year={2024},
  volume={72},
  number={10},
  month={Oct.},
  pages={6087-6101},
  doi={10.1109/TCOMM.2024.3394757}}

@book{nocedal2006numerical,
  author    = {Jorge Nocedal and Stephen J. Wright},
  title     = {Numerical Optimization},
  edition   = {2},
  publisher = {Springer},
  address   = {New York, NY, USA},
  year      = {2006}
}

@ARTICLE{Tzi2008VBI,
  author={Tzikas, Dimitris G. and Likas, Aristidis C. and Galatsanos, Nikolaos P.},
  journal={IEEE Signal Process Mag.}, 
  title={The variational approximation for {Bayesian} inference}, 
  year={2008},
  volume={25},
  number={6},
  month ={Nov.},
  pages={131-146},
  doi={10.1109/MSP.2008.929620}}

@ARTICLE{you20256G,
  author={You, Changsheng and Cai, Yunlong and Liu, Yuanwei and Di Renzo, Marco and Duman, Tolga M. and Yener, Aylin and Lee Swindlehurst, A.},
  journal={IEEE J. Sel. Areas Commun.}, 
  title={Next Generation Advanced Transceiver Technologies for 6{G} and Beyond}, 
  year={2025},
  volume={43},
  number={3},
  pages={582-627},
  month={Mar.},
  doi={10.1109/JSAC.2025.3536557}}

@ARTICLE{Zheng2022ris,
  author={Zheng, Beixiong and You, Changsheng and Mei, Weidong and Zhang, Rui},
  journal={IEEE Commun. Surveys Tuts.}, 
  title={A Survey on Channel Estimation and Practical Passive Beamforming Design for Intelligent Reflecting Surface Aided Wireless Communications}, 
  year={2022},
  volume={24},
  number={2},
  month={2nd Quart.},
  pages={1035-1071},
  doi={10.1109/COMST.2022.3155305}}

@ARTICLE{yuan2023k,
  author={Yuan, Zhiqiang and Zhang, Jianhua and Ji, Yilin and Pedersen, Gert Frølund and Fan, Wei},
  journal={IEEE Trans. Antennas Propag.	}, 
  title={Spatial Non-Stationary Near-Field Channel Modeling and Validation for Massive {MIMO} Systems}, 
  year={2023},
  volume={71},
  number={1},
  month={Jan.},
  pages={921-933},
  doi={10.1109/TAP.2022.3218759}}

@ARTICLE{Car2020nonstation,
  author={Carvalho, Elisabeth De and Ali, Anum and Amiri, Abolfazl and Angjelichinoski, Marko and Heath, Robert W.},
  journal={IEEE Wireless Commun.}, 
  title={Non-Stationarities in Extra-Large-Scale Massive {MIMO}}, 
  year={2020},
  volume={27},
  number={4},
  pages={74-80},
  month={Aug.},
  doi={10.1109/MWC.001.1900157}}

@INPROCEEDINGS{Mis2019onff,
  author={Mishra, Deepak and Johansson, Håkan},
  booktitle={Proc. IEEE Int. Conf. Acoust., Speech Signal Process. (ICASSP)}, 
  title={Channel Estimation and Low-complexity Beamforming Design for Passive Intelligent Surface Assisted {MISO} Wireless Energy Transfer}, 
  year={2019},
  volume={},
  number={},
  month={May},
  pages={4659-4663},
  doi={10.1109/ICASSP.2019.8683663}}

@INPROCEEDINGS{gao2023omp,
  author={Gao, Meng and Li, Huafu and Wang, Yang and Yang, Jiaao},
  booktitle={Proc. IEEE Int. Symp. Pers., Indoor Mobile Radio Commun. (PIMRC)}, 
  title={Sparsity Channel Estimation for Reconfigurable Intelligent Surface Aided {MIMO} Systems}, 
  year={2023},
  volume={},
  number={},
  pages={1-6},
  month={Sep.},
  doi={10.1109/PIMRC56721.2023.10293961}}

@ARTICLE{he2020matrix,
  author={He, Zhen-Qing and Yuan, Xiaojun},
  journal={IEEE Wireless Commun. Lett.}, 
  title={Cascaded Channel Estimation for Large Intelligent Metasurface Assisted Massive {MIMO}}, 
  year={2020},
  volume={9},
  number={2},
  pages={210-214},
  month={Feb.},
  doi={10.1109/LWC.2019.2948632}}

@ARTICLE{yang2023nearfield,
  author={Yang, Songjie and Lyu, Wanting and Hu, Zhenzhen and Zhang, Zhongpei and Yuen, Chau},
  journal={IEEE Trans. Veh. Technol.	}, 
  title={Channel Estimation for Near-Field {XL-RIS}-Aided {mmWave} Hybrid Beamforming Architectures}, 
  year={2023},
  volume={72},
  number={8},
  pages={11029-11034},
  month={Aug.},
  doi={10.1109/TVT.2023.3261340}}

@ARTICLE{dong2025doubleris,
  author={Dong, Erkang and Lian, Zhuxian and Wang, Yajun and Li, Yuanjiang and Su, Yinjie and Wang, Biao and Ling, Lin},
  journal={IEEE Internet Things J.	}, 
  title={Double-{IRS} Auxilary {mmWave} Near-Field Communications: Channel Modeling and Performance Analysis}, 
  year={2025},
  volume={12},
  number={11},
  month ={Jun.},
  pages={16023-16036},
  doi={10.1109/JIOT.2025.3530475}}

@inproceedings{Schniter2010Turbo,
  author    = {Philip Schniter},
  title     = {Turbo Reconstruction of Structured Sparse Signals},
  booktitle = {Proc. 2010 44th Annu. Conf. Information Sciences and Systems (CISS)},
  year      = {2010},
  pages     = {1-6},
  month     = mar,
  doi       = {10.1109/CISS.2010.5464920}
}

@INPROCEEDINGS{bian2024omp,
  author={Bian, Xuechun and Xu, Wenbo and Wang, Yue},
  booktitle={Proc. IEEE 100th Vehicular Technology Conference (VTC-Fall)}, 
  title={Sparse Representation-Based Robust Channel Estimation in {XL}-{RIS}-Assisted Systems}, 
  year={2024},
  volume={},
  month={Oct.},
  number={},
  pages={1-6},
  doi={10.1109/VTC2024-Fall63153.2024.10757853}}

@ARTICLE{lee2025rank,
  author={Lee, Jeongjae and Hong, Songnam},
  journal={IEEE Trans. Wireless Commun.}, 
  title={Near-Field {LoS}/{NLoS} Channel Estimation for {RIS}-Aided {MU-MIMO} Systems: Piece-Wise Low-Rank Approximation Approach}, 
  year={2025},
  month={Jun.},
  volume={24},
  number={6},
  pages={4781-4792},
  doi={10.1109/TWC.2025.3544198}}

@ARTICLE{ri2024ris,
  author={Rihan, Mohamed and Zappone, Alessio and Buzzi, Stefano and Fodor, Gabor and Debbah, Mérouane},
  journal={IEEE Network}, 
  title={Passive Versus Active Reconfigurable Intelligent Surfaces for Integrated Sensing and Communication: Challenges and Opportunities}, 
  year={2024},
  volume={38},
  month={May},
  number={3},
  pages={218-226},
  doi={10.1109/MNET.2023.3321542}}

@ARTICLE{shao2024ris,
  author={Shao, Xiaodan and You, Changsheng and Zhang, Rui},
  journal={IEEE Wireless Commun.}, 
  title={Intelligent Reflecting Surface Aided Wireless Sensing: Applications and Design Issues}, 
  year={2024},
  volume={31},
  month={Jun.},
  number={3},
  pages={383-389},
  doi={10.1109/MWC.004.2300058}}

@article{tuo2025nearfield,
  author  = {X. Tuo and Z. Chen and M.-M. Zhao and C. You and M.-J. Zhao},
  title   = {Near-Field Sparse {B}ayesian Channel Estimation and Tracking for {XL-IRS}-Aided Wideband {mmWave} Systems},
  journal = {arXiv preprint arXiv:2511.18752},
  year    = {2025},
}

@ARTICLE{wei2022ce,
  author={Wei, Yi and Zhao, Ming-Min and Zhao, Min-Jian and Cai, Yunlong},
  journal={IEEE Trans. Wireless Commun.}, 
  title={Channel Estimation for {IRS}-Aided Multiuser Communications With Reduced Error Propagation}, 
  year={2022},
  volume={21},
  number={4},
  month={Apr.},
  pages={2725-2741},
  doi={10.1109/TWC.2021.3115161}}

@INPROCEEDINGS{li2025lmmse,
  author={Li, Qingchao and El-Hajjar, Mohammed and Hemadeh, Ibrahim and Mestrah, Yasser and Shojaeifard, Arman and Hanzo, Lajos},
  booktitle={Proc. IEEE Int. Conf. Commun. (ICC)}, 
  title={Low-Complexity Channel Estimation for {RIS}-Assisted Multi-User Wireless Communications}, 
  year={2025},
  month={Jun.},
  volume={},
  number={},
  pages={6179-6184},
  doi={10.1109/ICC52391.2025.11160898}}

@INPROCEEDINGS{liu2021near,
  author={Liu, Kunzan and Zhang, Zijian and Dai, Linglong},
  booktitle={Proc. IEEE Global Communications Conference (GLOBECOM)}, 
  title={User-Side {RIS}: Realizing Large-Scale Array at User Side}, 
  year={2021},
  volume={},
  number={},
  pages={1-6},
  month={Dec.},
  doi={10.1109/GLOBECOM46510.2021.9685418}}

@ARTICLE{you2020ls,
  author={You, Changsheng and Zheng, Beixiong and Zhang, Rui},
  journal={IEEE J. Sel. Areas Commun.}, 
  title={Channel Estimation and Passive Beamforming for Intelligent Reflecting Surface: Discrete Phase Shift and Progressive Refinement}, 
  year={2020},
  volume={38},
  number={11},
  month={Nov.},
  pages={2604-2620},
  doi={10.1109/JSAC.2020.3007056}}
\end{document}